\documentclass[12pt,preprint]{aastex}

\slugcomment{Accepted to Ap.\ J.}

\shorttitle{Observation of ribbon substructure with IRIS}
\shortauthors{Brannon \& Longcope}

\begin{document}


\title{Spectroscopic observations of evolving flare ribbon substructure suggesting origin in current sheet waves}


\author{S.R. Brannon, D.W. Longcope, J. Qiu}
\affil{Department of Physics, Montana State University, Bozeman, MT 59717, USA}




\begin{abstract}
We present imaging and spectroscopic observations from the Interface Region Imaging Spectrograph (IRIS) of the evolution of the flare ribbon in the SOL2014-04-18T13:03 M-class flare event, at high spatial resolution and time cadence.  These observations reveal small-scale substructure within the ribbon, which manifests as coherent quasi-periodic oscillations in both position and Doppler velocities.  We consider various alternative explanations for these oscillations, including modulation of chromospheric evaporation flows.  Among these we find the best support for some form of wave localized to the coronal current sheet, such as a tearing mode or Kelvin-Helmholtz instability.
\end{abstract}



\section{Introduction}
\label{sec:intro}

The standard picture of the formation of flare ribbons in the solar chromosphere begins with magnetic reconnection in the corona \citep{kopp76}.  As magnetic field lines reconnect across a current sheet, coronal flare loops are formed.  The plasma along these loops is heated to flare temperatures both by the reconnection itself and by the subsequent contraction of the loop under magnetic tension \citep{longcope09}.  The energy from this flare plasma is then transported down the legs of the loop by non-thermal particles \citep{brown73}, thermal conduction \citep{craig76,forbes89}, wave propagation \citep{russell13}, or some other means, until it reaches the cool, dense chromospheric plasma located at the loop footpoints.  Energy is rapidly deposited in the chromosphere and transition region (TR), resulting in plasma flows both up and down the loop.  The upflows are called {\em chromospheric evaporation}, and are responsible for filling the flare loop with hot plasma.  Evaporative upflow of hot material is frequently accompanied by downward motions of cooler material, termed {\em chromospheric condensation} \citep{ichimoto84,brosius04,milligan09}.  Both evaporating and condensing components are significantly brighter than ambient material, and thus appear as elongated emission called flare ribbons.

As the flare progresses and new magnetic field lines are reconnected, the flare ribbons move outward from the polarity inversion line, which also marks the approximate location of the coronal current sheet \citep{kopp76}.  The flare ribbon emission is generally considered to represent the footpoints of recently reconnected field lines, and the ribbons are therefore an indirect image of the coronal reconnection process which is imprinted into the chromosphere \citep{forbes84}.  The structure and evolution of the ribbons can thus serve as an observational proxy for events occurring along the current sheet \citep{longcope07,qiu09,nishizuka09}.  Significant effort has been put into using observations of chromospheric ribbons to investigate the coronal reconnection \citep{schmieder87,falchi97,isobe05,miklenic07}, including the reconnection rate and current sheet electric field \citep{qiu02}.

The general trend of ribbon spreading during a flare has been established for several decades, however it has only been since the early 2000's that instruments with sufficient resolution and cadence, such as the {\em Transition Region and Coronal Explorer} ({\em TRACE}, Handy et al.\ 2009), have allowed for observations of small-scale substructure within the ribbon itself.  Several studies of {\em TRACE} observations found that flare ribbons in some events were broken into small and tightly-spaced bright sources, which were dubbed ``compact bright points'' (CBPs) \citep{warren01,fletcher03,fletcher04}.  In \cite{fletcher04}, CBPs were found to exhibit a random component to their motion in addition to the large-scale ribbon spreading, and they interpreted CBPs as individual loop footpoints that wander through the magnetic canopy during the course of the flare.  Additionally, they found that the brightness of a CBP was correlated to the product of footpoint speed and line-of-sight magnetic field, which they interpreted as a measure of the local coronal reconnection rate.

High-resolution observations of coronal loops, meanwhile, have strongly suggested that instabilities may occur within the reconnection region which subsequently induce oscillations in the flare loop \citep{aschwanden99,ofman11}.  One such instability which has been studied in the context of the solar corona is the {\em Kelvin-Helmholtz} instability, which occurs at a fluid interface with a discontinuity in flow speeds \citep{uchimoto91}.  KH instabilities have been specifically invoked to explain observed oscillations in coronal loops \citep{ofman11} and auroral spirals in the magnetopause \citep{lysak96}.  Another instability that has attracted interest is the {\em tearing mode} (TM) instability, in which magnetic islands spontaneously form and grow in a current sheet during reconnection \citep{furth63}.  The TM instability is believed to be important in permitting the fast reconnection rates required for solar flare energy release in so-called ``impulsive bursty reconnection'' models \citep{priest85}, and numerical simulations of reconnection have frequently invoked some form of velocity shear or TM instability along the current sheet \citep{karpen95,kliem00}.

Despite some successes \citep{aschwanden99,ofman11}, it has proven difficult to unambiguously resolve either current sheet instabilities or the associated flare loop oscillations, due in part to lower emission intensities and line-of-sight effects in the corona.  The chromospheric flare ribbons, on the other hand, are much more intense relative to the background emission, and any footpoint brightenings or ribbon motions are more easily distinguished from noise.  The connectivity between the reconnection region and the chromospheric footpoints means that a global instability in the coronal current sheet might be expected to imprint itself into the ribbon evolution.  Further, since the entire current sheet is presumably connected to the ribbons via the reconnecting flare loops, a current sheet instability would likely manifest as a substructure in the ribbons which could be easily distinguished from random CBP motion by a coherent pattern.  This assumes, of course, that field-aligned transport mechanisms translate energy consistently down the loops.  An alternative is that the observed ribbon patterns might be a result of the transport rather than the source.  The conventional understanding, based on numerous observations of coherent ribbon motion at two magnetically conjugate sites, is that the transport does not play a significant role.

In fact, the Interface Region Imaging Spectrograph (IRIS, De Pontieu et al.\ 2014), which makes high spatial and temporal resolution observations of the chromosphere and TR, has captured instances of such coherent substructure in flare ribbons.  One example is a recent study of an X-class flare by \cite{ting15}, where the authors identify a quasi-periodic slipping motion of flare loop footpoints which is observed as a series of small bright knots in Si \textsc{iv} which oscillate along the ribbon.  They interpret this slipping behavior as an apparent motion of footpoints, which brighten and fade as quasi-periodic slipping reconnection drives impulsive footpoint heating.  Finally, based on the timing of the oscillations, they suggest that the slipping reconnection may be due to varying densities in the current sheet which are driven by p-mode oscillations above the sunspots.

In this paper, we present an analysis of an IRIS observation from 2014 April 18 of a two-ribbon flare which displays coherent substructure of the ribbon during the impulsive phase of the flare, very similar to that reported by \cite{ting15}.  We find support in the data for a scenario of a flare loop undergoing elliptical oscillations, driven by either a TM or KH instability in the current sheet.  Our paper is outlined as follows: in Section \ref{sec:obs}, we describe the data and the details of the flare event.  Then, in Section \ref{sec:results}, we describe the evolution of the flare ribbons and our method for spectral line fitting, determine the Doppler velocities for Si \textsc{iv} within the ribbon, and lastly discuss both the behavior of other spectral lines and compare the substructure in both ribbons between conjugate points.  Next, in Section \ref{sec:interpretation}, we consider and dismiss two alternative scenarios for generating the observed ribbon substructure before proposing our scenario for instability-driven elliptical oscillation.  Finally, in Section \ref{sec:discussion}, we discuss some of the possible implications for our results.

\section{Observation}
\label{sec:obs}

\subsection{Instrument}
The Interface Region Imaging Spectrograph (IRIS) is a space-based observatory in low-Earth orbit that was launched on 2013 June 27.  The primary instrument onboard IRIS is a dual-range UV spectrograph (SG) with $0''.16$ pixels and an effective effective spatial resolution of $0''.4$.  The SG slit is $0''.33$ wide and $175''$ long, and covers FUV passbands from 1332 \AA\ to 1358 \AA\ and 1389 \AA\ to 1407 \AA\ and an NUV passband from 2783 \AA\ to 2835 \AA. These passbands include lines formed over a wide range of temperatures from the photosphere (5000 K) to the corona (1 to 10 million K).  Two fiducials (dark bands where light is excluded from the SG) are located along the slit which provide alignment and spatial context, and the entire SG can be rastered back-and-forth to provide coverage over a 2-D area \citep{iris14}.

The other instrument featured onboard IRIS is a slit-jaw imager (SJI) which records context images of the observation region on either side of the SG.  The IRIS SJI includes four wavelength passbands, including two transition region lines (Si \textsc{iv} 1400 \AA\ and C \textsc{ii} 1335 \AA), with a field-of-view (FOV) of $175''$ $\times$ $175''$, and typically operates at one-third the cadence of the SG (due to cycling through the passbands) \citep{iris14}. The inclusion of a SJI, with the same FOV, resolution, and alignment as the SG allows for detailed contextual knowledge of the slit placement relative to observational features such as sunspots, ribbons, and coronal loops.  It also facilitates very precise co-alignment ($<1''$) between IRIS and other instruments, such as SDO/AIA, by aligning visual features in the SJI to images from those instruments.

\subsection{Flare}
\begin{figure}
\epsscale{1.0}
\plotone{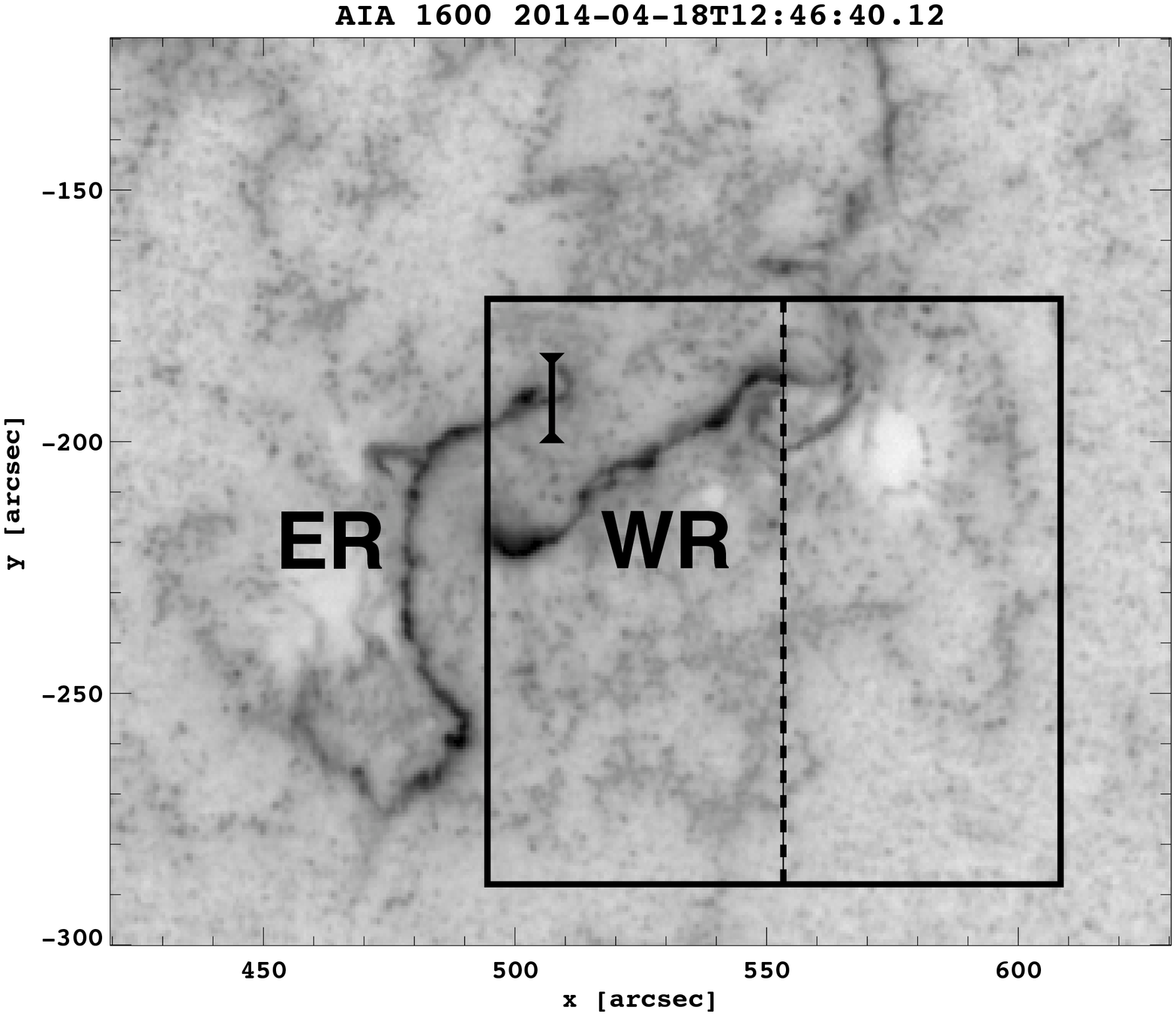}
\caption{SDO/AIA 1600 \AA\ image of the flare ribbons, in reversed $\log_{10}$ black-white (RLBW).  The black inset box is the positioning of the IRIS SJI FOV, and the vertical black dashed line is the IRIS SG slit position.  The east and west ribbons are indicated by ``ER'' and ``WR'', respectively.  The black ``I''-shaped line is an artificial slit discussed in Section \ref{sec:east}.
\label{fig:aia1600}}
\end{figure}

The event investigated in this paper is a GOES M7.3-class flare that occurred on 2014 April 18.  The flare was located in NOAA Active Region 12036, which on that day was located at approximately $500''$ West and $200''$ South from solar disk center.  The flare start time was 12:31 UT and the GOES X-ray flux peaked at approximately 13:03 UT, with the event mostly concluded by 13:20 UT.  The large-scale structure consists of two chromospheric ribbons, one on the east side and one on the west side of the flare, visible as strands of intense 1600 \AA\ emission at 12:46:40 UT in Figure \ref{fig:aia1600} (note that the color scale is a reversed $\log_{10}$ black-white (RLBW), with white as lowest intensity and black as highest intensity).    We have labeled the east and west ribbons accordingly as ``ER'' and ``WR'', and we refer to the ribbons as such throughout the text.  Both ribbons generally follow a single distinct path from southeast to northwest, although the ER is more strongly tilted toward a north-south orientation.  There are also occasional offshoots and patchy regions of intensity which evolve with time.

\begin{figure}
\epsscale{1.0}
\plotone{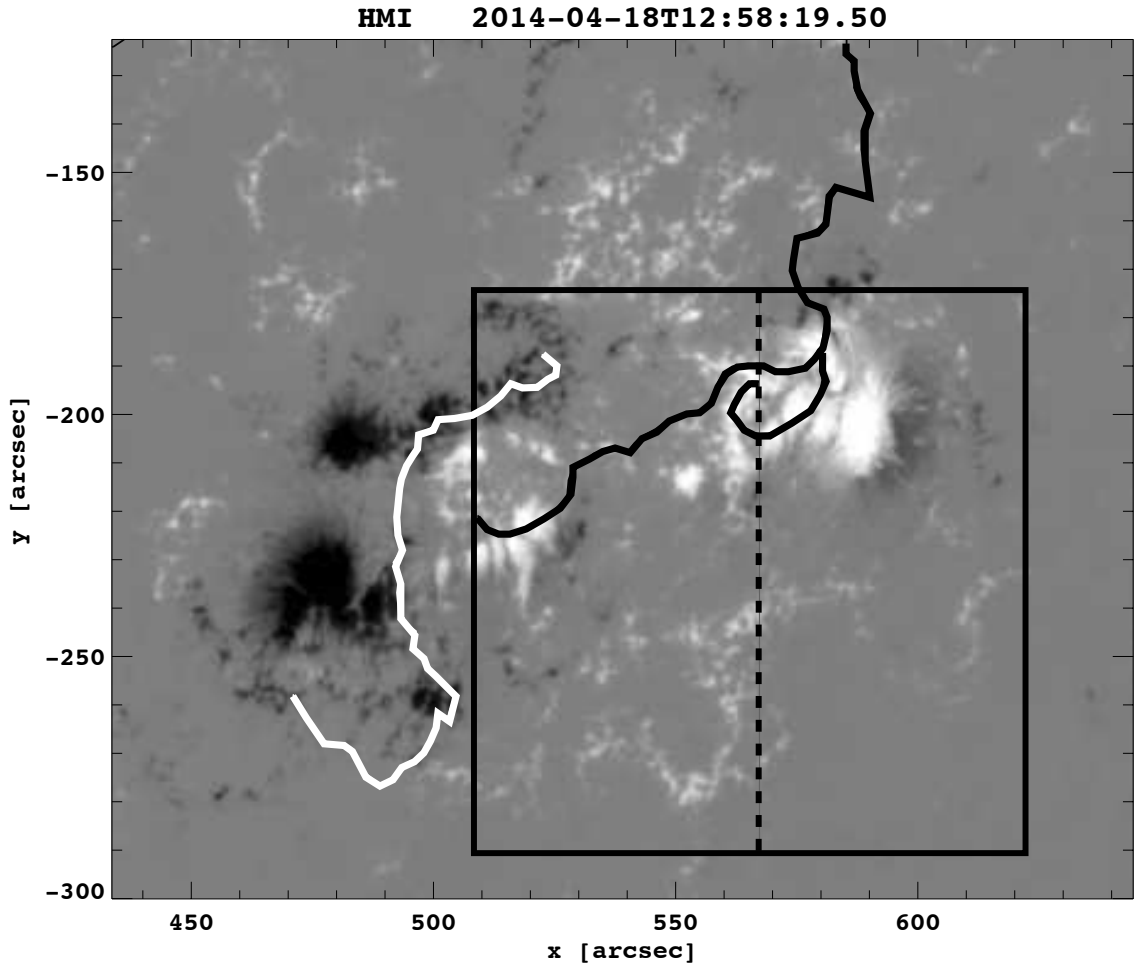}
\caption{SDO/HMI magnetogram of the flare active region, with standard colors.  The balck inset box is the positioning of the IRIS SJI FOV, and the vertical black dashed line is the IRIS SG slit position.  The black and white solid lines trace the west (WR) and east (ER) ribbons from Figure \ref{fig:aia1600}.
\label{fig:hmi}}
\end{figure}

One notable exception to the general SE-to-NW single-track ribbon paths can be seen in the WR starting at around $550''$ West and $200''$ South (Figure \ref{fig:aia1600}).  At this point, moving SE to NW, the ribbon clearly splits into two channels which merge back together farther to the NW, resulting in an isolated island between the ribbon channels.  This split ribbon feature persists over most of the evolution of the flare and appears in both SDO/AIA 1600 \AA\ and IRIS SJI 1400 \AA, but does not appear in the SDO/AIA 171 \AA\ passband.  The most likely origin of this feature can be seen in the SDO/HMI magnetogram for this region taken in the middle of the flare duration (12:58:19 UT) as shown in Figure \ref{fig:hmi}, where positive and negative LOS polarity are colored white and black, respectively, and the ER and WR positions are traced for context as the white and black lines.  In the center of the split WR island we observe a significant neutral gap running diagonally through the positive magnetic region centered at $575''$ West and $200''$ South.  The split 1600 \AA\ ribbon appears to track around either side of this neutral channel.

There is also a prominent arcade of coronal loops which appears in the SDO/AIA 171 \AA\ passband approximately 30 minutes after the first appearance of the flare ribbons.   Figure \ref{fig:aia171} (also RLBW) shows a 171 \AA\ image of the flare region taken at 13:12:37 UT showing the flare ribbons shortly after they first appear.  For comparison with Figure \ref{fig:aia1600} we have traced the position of the ribbons as jagged white lines.  The loops visible in Figure \ref{fig:aia171} all follow roughly E to W paths between the two ribbons, although the exact loop connections between the ER and WR are not easy to establish because of the time delay for the appearance of the loops in this passband.  We also searched for loops at earlier times using higher temperature SDO/AIA passbands, such as 131 \AA\ and 193 \AA, but saturation and lack of sharpness in these passbands relative to 171 \AA\ impedes clear identification.  Our main use for the AIA imaging data is to identify magnetic conjugacy.  We use 171 \AA\ images for this because they are sharpest and will probably extend to the lowest points along the loops.  The 171 \AA\ image clearly indicates a dipolar component to the magnetic geometry, which we confirm with the SDO/HMI magnetogram in Figure \ref{fig:hmi}.  We clearly see two dominant polarities, negative to the east and positive to the west, which the ribbons track quite closely and which are connected by the flare loops.  Finally, we note there was a coronal mass ejection (CME) associated with this flare (observed by LASCO); the flux rope eruption responsible for the CME is discussed in more detail in \cite{cheng15}.

The IRIS SG data we use for this study is drawn from a sit-and-stare observation (i.e.\ no rastering of the SG slit) which began at 12:33:38 UT, shortly after the GOES start time for the flare, and which continued until 17:18:10 UT.  This observation thus covers nearly all of the impulsive rise phase and the entire decay phase of the flare.  The SG cadence for this observation was $\sim$9 s, while the SJI \AA\ cadence for 1400 \AA\ was $\sim$27 s, and the SJI FOV was initially centered at $550''$ West, $230''$ South (note that the spatial metric at this position on the disk is $\sim$900 km/arcsec).  An example SJI 1400 \AA\ image, taken at 12:46:34 UT (approximately the same time as Figure \ref{fig:aia1600}), is shown in Figure \ref{fig:iris1400} (we again use RLBW).  For context, note that the IRIS SJI FOV is shown as a black inset box in Figures \ref{fig:aia1600}-\ref{fig:hmi}, and that the IRIS SG slit position is indicated in those three figures by a vertical dashed black line.  Finally, note that the IRIS SG slit is conveniently positioned across the WR, directly over the location of the magnetic gap and ribbon island mentioned above.  In summary, the slit placement, high cadence, and consistently stable rotation tracking during this observation make it ideal for a spectroscopic study of the fine structure of flare ribbons and its evolution.

\subsection{Wavelength correction}
The rest wavelengths of spectral lines for the IRIS SG are generally dynamically shifted to some degree due to the orbital motion and thermal variations of the IRIS spacecraft and must be corrected for every spectrograph exposure.  To do this we use the SSWIDL routine \verb+iris_orbitvar_corr_l2+, which fits the Ni \textsc{i} 2799.474 \AA\ line to determine the appropriate wavelength shifts for the FUV bandpass \citep{iris14}.  Even after applying this correction, however, we find that the peak of the O \textsc{i} 1355.598 \AA\ is systematically redshifted during the observation by $\sim$0.01 \AA, even though as a photospheric line it should generally be stationary. We therefore subtract this additional shift from the FUV wavelength axis to make the O \textsc{i} line stationary, and subsequently reference all other lines to their CHIANTI database line centers \citep{itn20}.

\section{Results}
\label{sec:results}

\subsection{Ribbon evolution}
\label{sec:evolution}
\begin{figure}
\epsscale{1.0}
\plotone{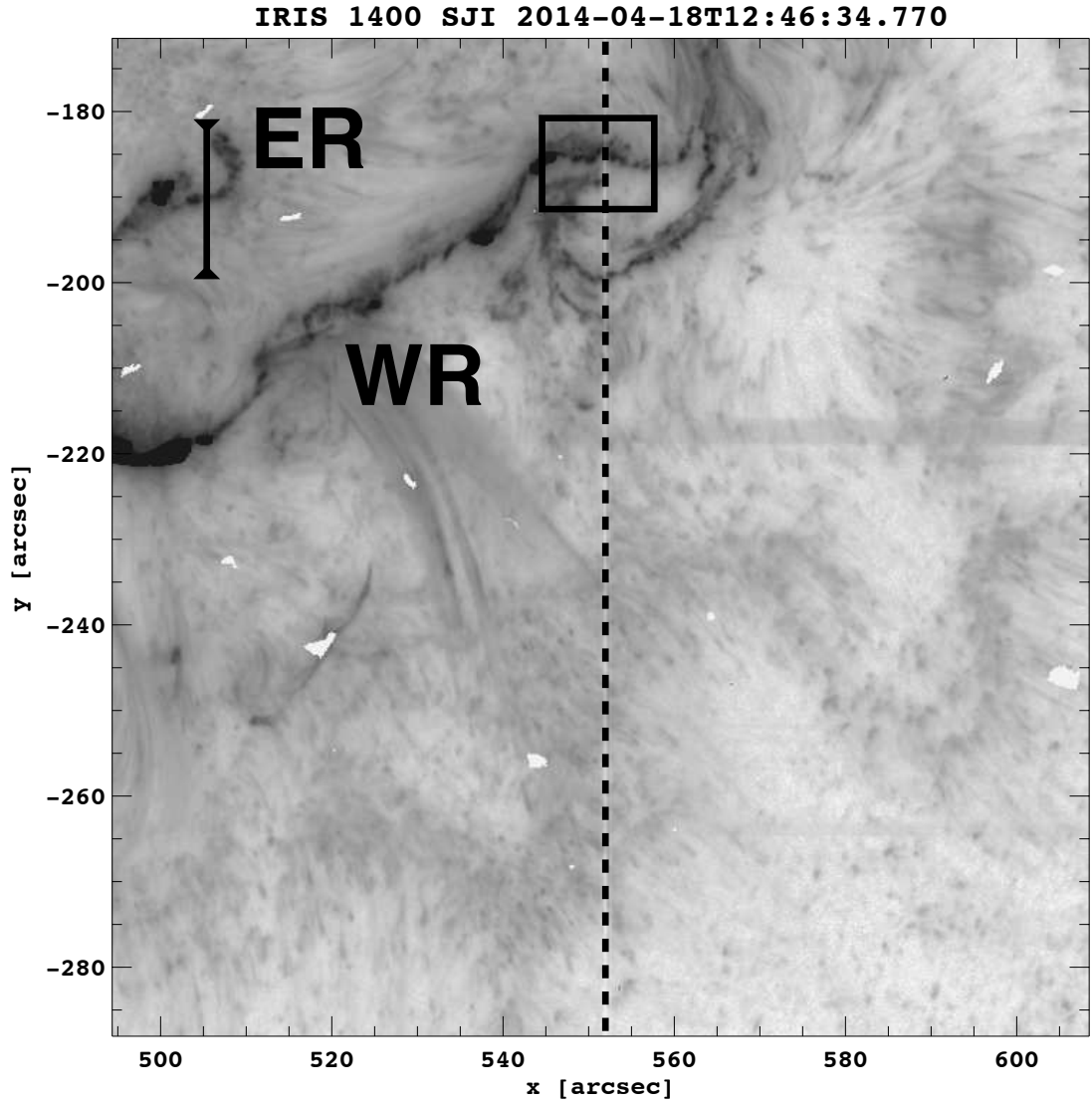}
\caption{An image of the flare ribbons from the IRIS 1400 \AA\ SJI, in RLBW, corresponding to the black inset boxes for Figures \ref{fig:aia1600}-\ref{fig:hmi}.  The east and west ribbons are indicated by ``ER'' and ``WR'', respectively.  The vertical black dashed line is the IRIS SG slit, and the black inset box outlines where the WR crosses the slit and the positioning of the frames for Figure \ref{fig:ribbon_movie}.  The black ``I''-shaped line is an artificial slit discussed in Section \ref{sec:east}.
\label{fig:iris1400}}
\end{figure}

\begin{figure}
\epsscale{1.0}
\plotone{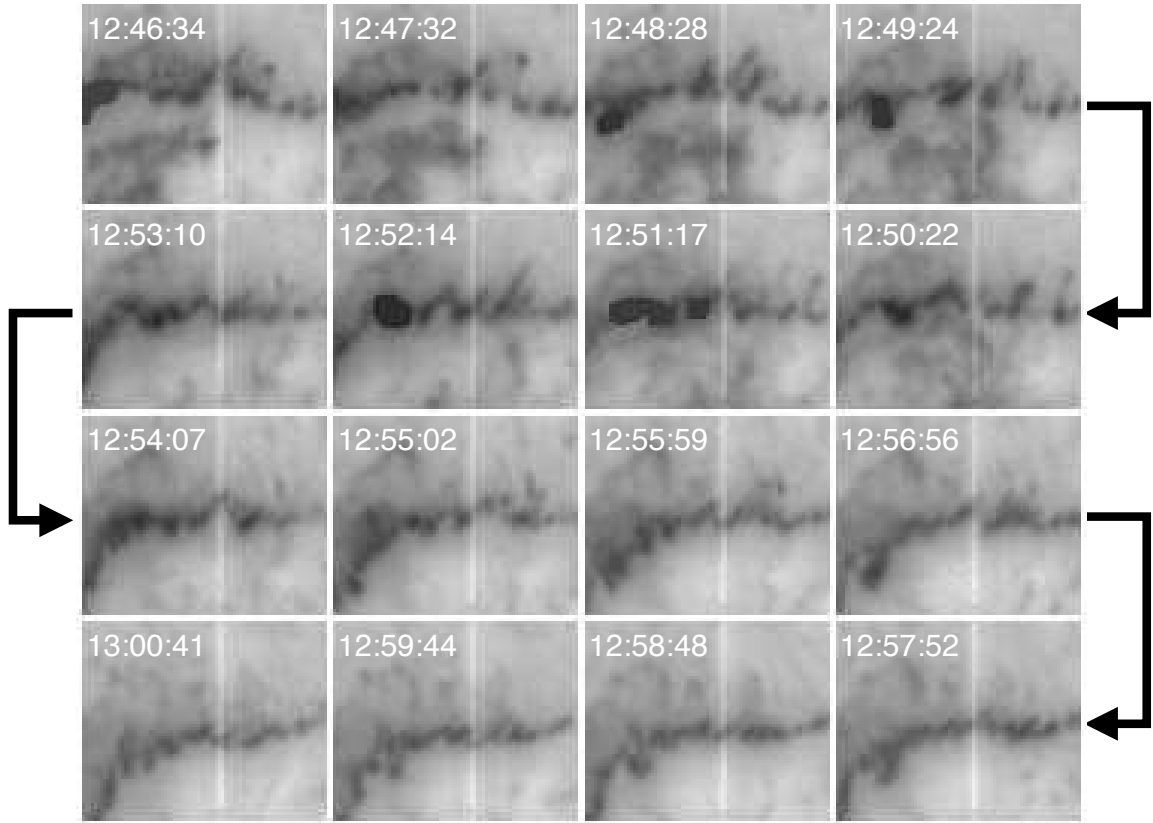}
\caption{Time series of SJI 1400 \AA\ images, for the inset box in Figure \ref{fig:iris1400}, showing the evolution of the flare ribbon in $\sim$1 minute intervals beginning at 12:46:34 UT, in RLBW.  Time for each image is displayed at upper left for each frame.  Note that time runs from left-to-right in the first row, then right-to-left in the second row, and continues to alternate down the rows, as indicated by the arrows.
\label{fig:ribbon_movie}}
\end{figure}

The flare ribbon appears as an intense band in the northern portion of the IRIS SJI beginning around 12:46:34 UT.  The overall orientation of the WR is from SE to NW, and in the vicinity of the SG slit the ribbon is running very nearly due E-W, across and perpendicular to the slit.  As the WR evolves it slowly drifts S, generally maintaining its perpendicular orientation to the slit.  This motion is away from the polarity inversion line (PIL), in agreement with the classic picture of ribbon spreading \citep{kopp76}.  Figure \ref{fig:ribbon_movie} shows a series of 16 images from the SJI 1400 \AA\ passband, taken from the black inset box near the top of Figure \ref{fig:iris1400}, at $\sim$1 minutes intervals starting at 12:46:34 UT (the same time as Figure \ref{fig:iris1400}) and continuing to 13:00:41 UT.  The color scale is RLBW.  The inset frame has been chosen to cover the area of the SJI where the northern branch of the WR crosses the SG slit.  The reader should note that the time axis for these 16 frames runs from left-to-right in the top row, then right-to-left in the second row, and so on row-to-row as indicated by the arrows to form a movie of the ribbon evolution.  The SG slit can be seen as a pale line running down the middle of each frame, with the upper fiducial just visible at the bottom of each frame.  Intermittent saturated pixels appear to the left of the slit in some frames (for example, 12:52:41 UT).

In addition to the large-scale evolution of the ribbon shown in Figure \ref{fig:ribbon_movie}, we also note a distinct substructure that evolves on a smaller scale and at a faster rate than the overall ribbon motion.  This substructure appears in most frames as a jagged sawtooth pattern that cuts across the slit, with multiple patches of bright emission oriented diagonally to the slit that break up the ribbon.  A very similar structure was observed by \cite{ting15} and called ``slipping reconnection'' by them.   Since this term also refers to a model \citep{aulanier06}, which may or may not be pertinent to the observation, we use the term sawtooth here instead.  Times at which this pattern appears most prominent include 12:50:49 UT, 12:52:14 UT, and 12:55:32 UT.  An inspection of this sawtooth pattern from frame-to-frame reveals that the diagonal features appear to slide across the slit from east to west as the ribbon drifts south over time.  Our best estimate of its pattern speed is $v_{st}\approx 15$ km s$^{-1}$, parallel to the ribbon itself.  One example where this is most readily apparent occurs between 12:53:38 and 12:55:32 UT.  In these three frames a diagonal sawtooth begins just to the left of the slit, moves west such that one minute later it is crossing the slit, and one minute after that is seen on the right side of the slit.

\begin{figure}
\epsscale{1.0}
\plotone{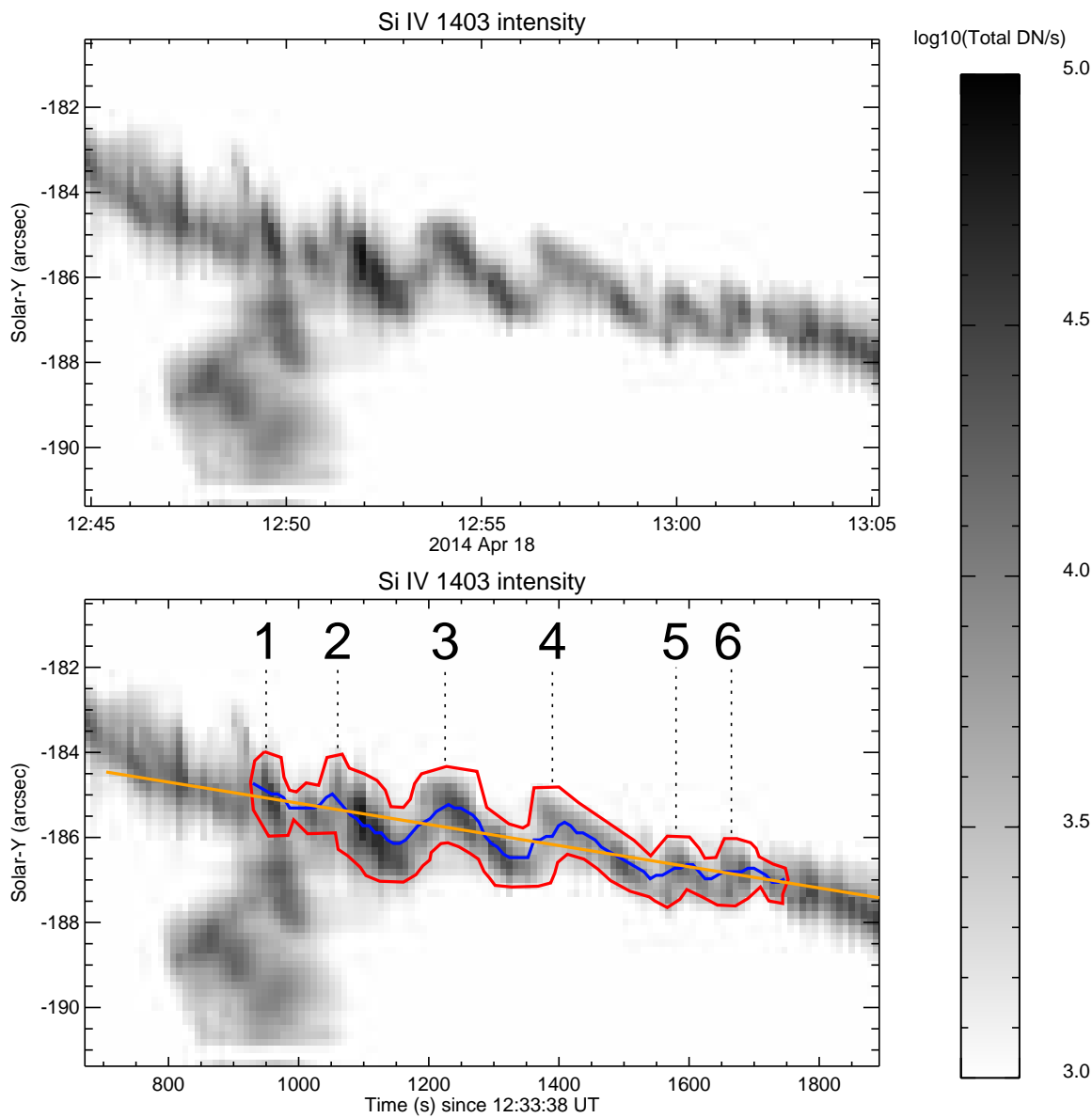}
\caption{Upper panel: time-distance stackplot of the total Si \textsc{iv} 1403 \AA\ SG passband intensity, in RLBW.  Time is given on the $x$-axis in UT, and the $y$-axis is solar-Y in arcsec. Intensity scale is given to the right.  Lower panel: reprint of the upper panel with a red outline indicating the position of the sawtooth pattern described in Section \ref{sec:evolution}.  Time is given on the $x$-axis in s (after 12:33:38 UT), and the $y$-axis is unchanged.  The blue line indicates the sawtooth centroid position, and the orange line is a linear fit to the blue line.  The numbers 1-6 indicate six peaks in the sawtooth oscillation.
\label{fig:intensity_stack}}
\end{figure}

Since the sawtooth pattern in the SJI 1400 \AA\ is moving across the SG slit, we would expect the pattern to also be reflected in spectral intensity for the Si \textsc{iv} 1403 \AA\ passband.  To calculate the total intensity in this passband, we first divide the data for each SG frame by the appropriate exposure time to obtain normalized units of DN s$^{-1}$.  We then reset all negative values in the passband (which occur near the passband edges) to 0, and finally sum the data over the entire Si \textsc{iv} 1403 \AA\ passband.  In the upper plot of Figure \ref{fig:intensity_stack} we show the resulting time-distance stackplot for the time range 12:45 UT to 13:05 UT.  The $y$-axis shows the heliographic position of the slit pixels at at the beginning of the observation; there is some wobble to $y$-pointing of the instrument, but it is less than 0.3$''$ over the entire observation.  Data values vary between 20 DN s$^{-1}$ and 64000 DN s$^{-1}$.  We again note the southward motion of the ribbon, beginning at $\sim$$184''$ S and subsequently moving south to $\sim$$188''$ S.  We also observe immediately that the sawtooth pattern is indeed present in the SG data starting around 12:48 UT and ending at around 13:02 UT, and that three teeth in particular can be clearly seen between 12:51 UT and 13:01 UT.  There is also a feature that moves up from the south beginning at around 12:47 UT, however this is a piece of the northward-moving southern branch and not part of the ribbon sawtooth structure.

In the lower plot of Figure \ref{fig:intensity_stack}, we have replotted the intensity stackplot (with the time axis in seconds this time), and we have overlaid a red contour line around the section of the ribbon emission for which the sawtooth pattern is the clearest.  Henceforth, when referring to the ``sawtooth'', we will be referring to this outlined region of pixels.  The numbers 1-6 indicate six peaks in the sawtooth oscillation, with dashed lines also indicating the approximate beginning time of a southward motion in the sawtooth.  The oscillation period varies between 80 and 190 s, averaging $\sim$140 s, consistent with sawtooth structures $\sim$2 Mm long moving across the slit at $\sim$15 km s$^{-1}$. We have also overlaid the sawtooth centroid position (as a function of time) as the blue line, as well as a linear best fit to the sawtooth mean as the orange line.  The linear fit has a southward velocity of 1.6 km/sec, while the mean position moves at velocities ranging from  $\pm20$ km/s.  This confirms the much faster motion of the sawtooth substructure relative to the overall ribbon motion.  Finally, we observe that the shallow sides of the sawtooth pattern are all directed away from the PIL located between the ER and the WR.

\subsection{Spectral line fitting}
\label{sec:fitting}
\begin{figure}
\epsscale{1.0}
\plotone{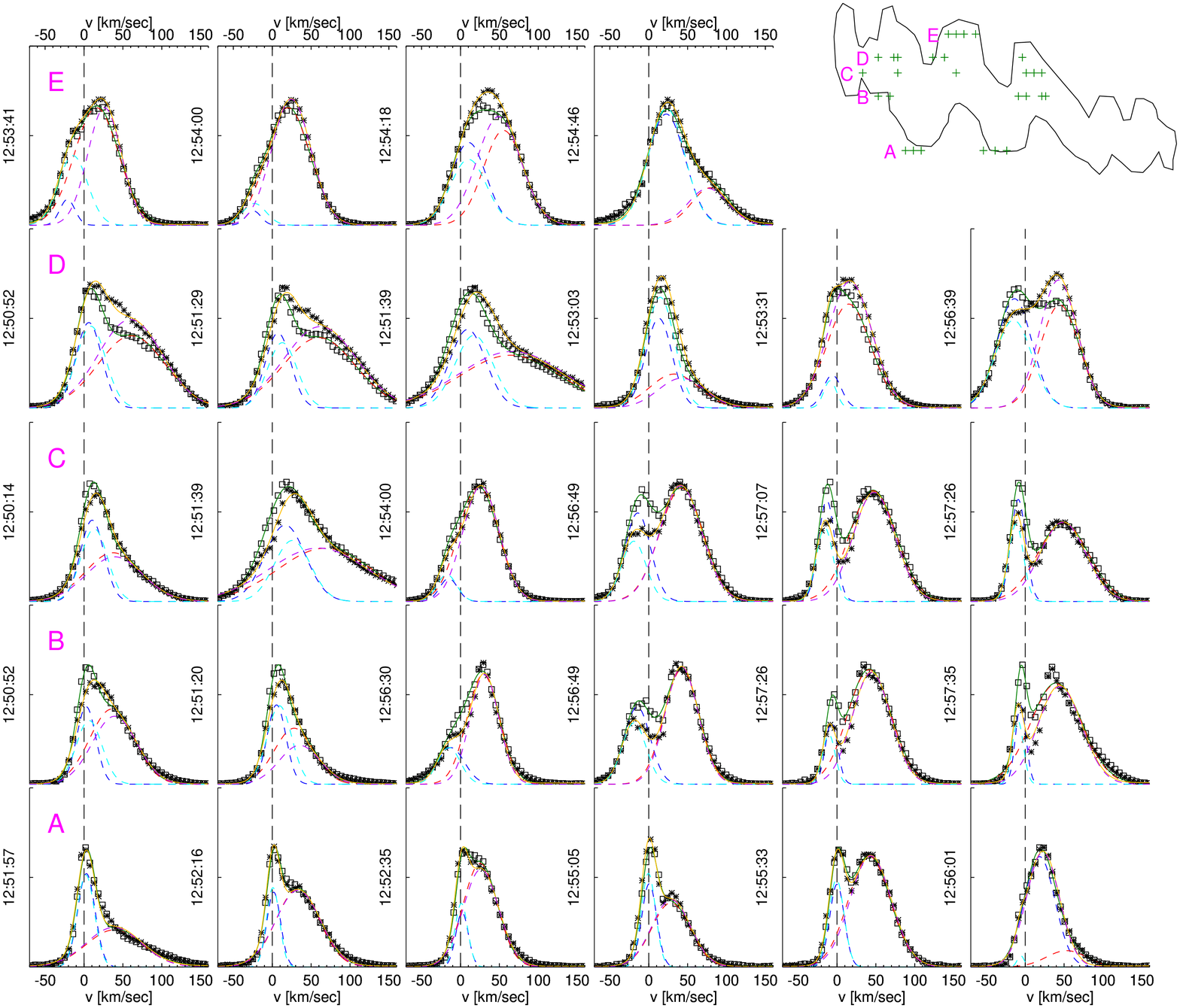}
\caption{Selected Si \textsc{iv} spectral lines for 28 pixels in or near the sawtooth pattern shown in Figure \ref{fig:intensity_stack}.  An outline of the sawtooth is shown at upper right, and the five rows of plots ``A''-``E'' each sample a given spatial position at several different times, as indicated by the ``+'' symbols at upper right.  Each plot displays the spectrograph data for Si \textsc{iv} 1394 \AA\ (as asterisks) and 1403 \AA\ (as squares), with the time printed to the right of each plot.  The $x$-axis is in km s$^{-1}$ from nominal line center (same for both lines), and the $y$-axis scaling is arbitrary (the 1394 \AA\ data has been scaled by a factor of 0.5 for direct comparison to the 1403 \AA\ line).  Additionally, the solid orange (green) lines are the two component Gaussian fits for the 1394 \AA\ (1403 \AA ) spectral line, as described in Section \ref{sec:fitting}.  The dashed lines are the individual components $C_B$ and $C_R$ for each of the Si \textsc{iv} spectral lines, with cyan/violet used for the respective 1394 \AA\ spectral components and blue/red used for the respective 1403 \AA\ spectral components.
\label{fig:si4_compare}}
\end{figure}

The WR sawtooth substructure is seen in the 1400 \AA\ SJI passband.  These images are dominated by the two Si \textsc{iv} lines (1394 \AA\ and 1403 \AA) which are recorded in the IRIS SG FUV 2 range \citep{iris14}. These two lines are ideal for examining flare ribbon plasma, since at $\sim$80,000 K they are located within the transition region at the footpoints of the flaring loops.  Additionally, they are typically strong lines that dominate their respective spectral regions and therefore are easily distinguished from the lines of other ion species.  In Figure \ref{fig:si4_compare} we have plotted the exposure-normalized data for the two lines for several representative positions and times in and around the sawtooth pattern: 1394 \AA\ as asterisks and 1403 \AA\ as squares.  The $x$-axes are in units of km s$^{-1}$ with $v=0$ referring to the rest wavelength defined as described above, where redshifts and blueshifts are defined as positive and negative velocities respectively.  The $y$-axis is arbitrarily scaled for each pixel, and the time appears to the left of each plot.  At the upper right is the outline of the sawtooth from Figure \ref{fig:intensity_stack}, with ``+'' symbols arranged in five rows ``A''-``E'', corresponding to a row of spectral plots, which indicate the position and time of each spectral plot.  Finally, note that the 1394 \AA\ data have all been scaled by a factor of 0.5, so that the two spectral lines will appear on the same scale (and see the final paragraph of this section for more detail on this choice).

The positions and times (henceforth ``pixels'') chosen for Figure \ref{fig:si4_compare} have been chosen to illustrate several general features of the Si \textsc{iv} spectral lines in this event.  First, the majority of the pixels within the sawtooth appear to contain two Gaussian components, as first noted by \cite{cheng15} in the early ribbon development.  The two components are present for both the 1394 \AA\ and the 1403 \AA\ lines, clearly demonstrating that both components are from Si \textsc{iv} and not from an accidental blend with another spectral line.  A fraction of pixels late in the sawtooth contain one or more additional components; however, these occur only in a minority of cases and typically have small amplitude compared to the two dominant components.  The lack of additional components, or of significant non-Gaussian tails in the spectral lines, suggests that the IRIS SG is not observing many separate loop footpoints within a single pixel, but rather only one or two distinct footpoints.  We also observe that each component appears to persist and to evolve within its row.  Both components are Doppler-shifted, with one component consistently redshifted and the other switching between redshift and blueshift, and both the Doppler shift and the velocity separation between the components evolves with time.  Finally, we note that neither component consistently dominates, and for some pixels they have comparable magnitude.

Within the sawtooth, there are numerous cases where at least one, and sometimes both, of the Si \textsc{iv} lines saturate the IRIS SG pixels.  This is true for $\sim$17\% of pixels for 1394 \AA\ and $\sim$5\% of pixels for 1403 \AA, and makes accurately fitting a two-component Gaussian problematic.  Except for saturation, however, most positions display behavior similar to that discussed above, although we discuss some additional discrepancies below.  We therefore assume that a two-component Gaussian will be appropriate for the majority of the pixels within the sawtooth.

Since there are too many positions and times within the sawtooth to manually supervise the fitting, we have developed an automated procedure for eliminating poor fits.  We note first that the chi-squared measure for the fits were found not to neatly discriminate between fits which were deemed good and bad by visual inspection.  Therefore, in our procedure the two-component fit is rejected if any one of the following criteria are met for either component:
\begin{enumerate}
\item any pixel in the band is greater than 2000 DN s$^{-1}$ (approximate saturation limit);
\item the Gaussian amplitude is less than 10 DN s$^{-1}$;
\item the line center is greater than $\pm$0.625 \AA\ ($\sim$134 km s$^{-1}$) from the rest wavelength;
\item the Gaussian width is greater than 0.625 \AA.
\end{enumerate}
We find that $\sim$21\% of 1394 \AA\ and $\sim$7\% of 1403 \AA\ attempted fits are rejected using this method; roughly two-thirds of these rejections are due to condition 1, while the rest are from the other three conditions.  Because of the greater percentage of successful fits for 1403 \AA, we will use only that spectral line for the remainder of the analysis starting in Section \ref{sec:doppler}.  In a small number ($<$1\%) of cases, only one component is the cause of the rejected fit; however, due to the rarity of these cases, we simply reject both components.  Finally, the two components are sorted according to their Doppler shifts, and we label the bluer component $C_B$ and the redder component $C_R$.

In Figure \ref{fig:si4_compare} we have plotted the resulting two-component Gaussian fits for the 28 selected pixels over the data for those pixels, for both the 1394 \AA\ (as the solid orange curve) and the 1403 \AA\ (as the solid green curve) spectral lines.  The dashed lines correspond to the individual components $C_B$ and $C_R$ for each of the Si \textsc{iv} spectral lines, with cyan/violet used for the respective 1394 \AA\ spectral components and blue/red used for the respective 1403 \AA\ spectral components.  We note that our criteria are generally adequate to reject bad fits, and the fits that remain conform to the data for both spectral lines quite well.  We also note that the fit parameters of the two components are generally similar for both spectral lines, which adds confidence that the fitting procedure is finding the same two major components.  We find that $C_B$ is generally narrower than $C_R$ (for $\sim$88\% of pixels).  The average non-thermal widths, given by $\sigma_{{\scriptsize \textrm{nt}}} = \sqrt{\sigma^2 - \sigma_{{\scriptsize \textrm{th}}}^2 - \sigma_{{\scriptsize \textrm{inst}}}}$ where $\sigma_{{\scriptsize \textrm{th}}} = 6.86$ km s$^{-1}$ and $\sigma_{{\scriptsize \textrm{inst}}} = 3.9$ km s$^{-1}$ \citep{depontieu15}, for the two components are $\sigma_{{\scriptsize \textrm{nt}}}^{{\scriptsize \textrm{B}}} = 19$ km s$^{-1}$ and $\sigma_{{\scriptsize \textrm{nt}}}^{{\scriptsize \textrm{R}}} = 38$ km s$^{-1}$.  We note that the non-thermal widths for $C_B$ are quite consistent with those found in quiescent active regions \citep{depontieu15} whereas $C_R$ shows much greater broadening, but we will defer discussion of this fact until later.

In an optically-thin plasma without geometrical effects, the 1394 \AA\ and 1403 \AA\ Si \textsc{iv} lines will have a 2:1 intensity ratio \citep{mathioudakis99}.  The majority of the sawtooth pixels display this 2:1 ratio: e.g.\ the majority of the scaled data for rows A, D, and E all lie on top of each other.  Some pixels, however, depart from the strict 2:1 line ratio, suggesting some optical depth effects.  This departure is more common in $C_B$, especially when it is very narrow, as in rows B and C.  It can also occur in $C_R$, as it does in row D.  In some cases the line ratio is less than two, while in other it is greater (as in $C_R$ of row D).  The former situation is commonly attributed to loss from the thicker line (1394 \AA) by scattering out of the line of sight \citep{mathioudakis99}.  The latter can occur in more complicated geometry when photons are scattered {\em into} the line of sight \citep{kerr05}.  The fact that we observe both cases in neighboring pixels suggests, not surprisingly, that the emitting plasma has a complex geometrical structure over scales comparable to the IRIS resolution.  Almost all optical-depths effects we observe are similar to those shown in rows B, C and D.  The line ratio remains very close to the optically thin value of 2:1, and the thicker component (1394 \AA) is still well fit by a Gaussian.  We therefore conclude that Si \textsc{iv} is never far from being optically thin and that Doppler shifts and line width are well measured by the Gaussian fit.

\subsection{Doppler velocities}
\label{sec:doppler}
In Figure \ref{fig:doppler_stack}, we plot the Doppler velocities determined above for both components using a blue-red color scale on time-distance stackplots ($C_B$ in the upper panel and $C_R$ in the lower panel).  The time axes follow the same labeling scheme as in Figure \ref{fig:intensity_stack}, and the blue-red color range runs between $\pm$40 km s$^{-1}$ for $C_B$ and between $\pm$80 km s$^{-1}$ for $C_R$ (in this system, we define redshifts as positive and blueshifts as negative).  We have also outlined the sawtooth pattern from Figure \ref{fig:intensity_stack} with the solid black line to help guide the reader in determining the relative position of features.  Finally, we have color-coded any pixels not fit (for any reason) by the above automatic routine as gray.

There are several important features to note in the Doppler velocity data, mostly for component $C_B$.  The first is that, as mentioned above, the Doppler shifts for component $C_B$ are sometimes red and sometimes blue.  However, we now observe that these shifts are not random, but are clustered into large regions of redshift and blueshift.  Additionally, we note that the blueshifts tend to be concentrated near the upswings and peaks in the sawtooth pattern, whereas the redshifts tend to be on the downswings and troughs (particularly prominent in the teeth marked ``3'' and ``4'').  Although it is not as apparent, this same pattern is reflected in component $C_R$, except that the redshift is merely weakened instead of shifted to the blue.

\begin{figure}
\epsscale{1.0}
\plotone{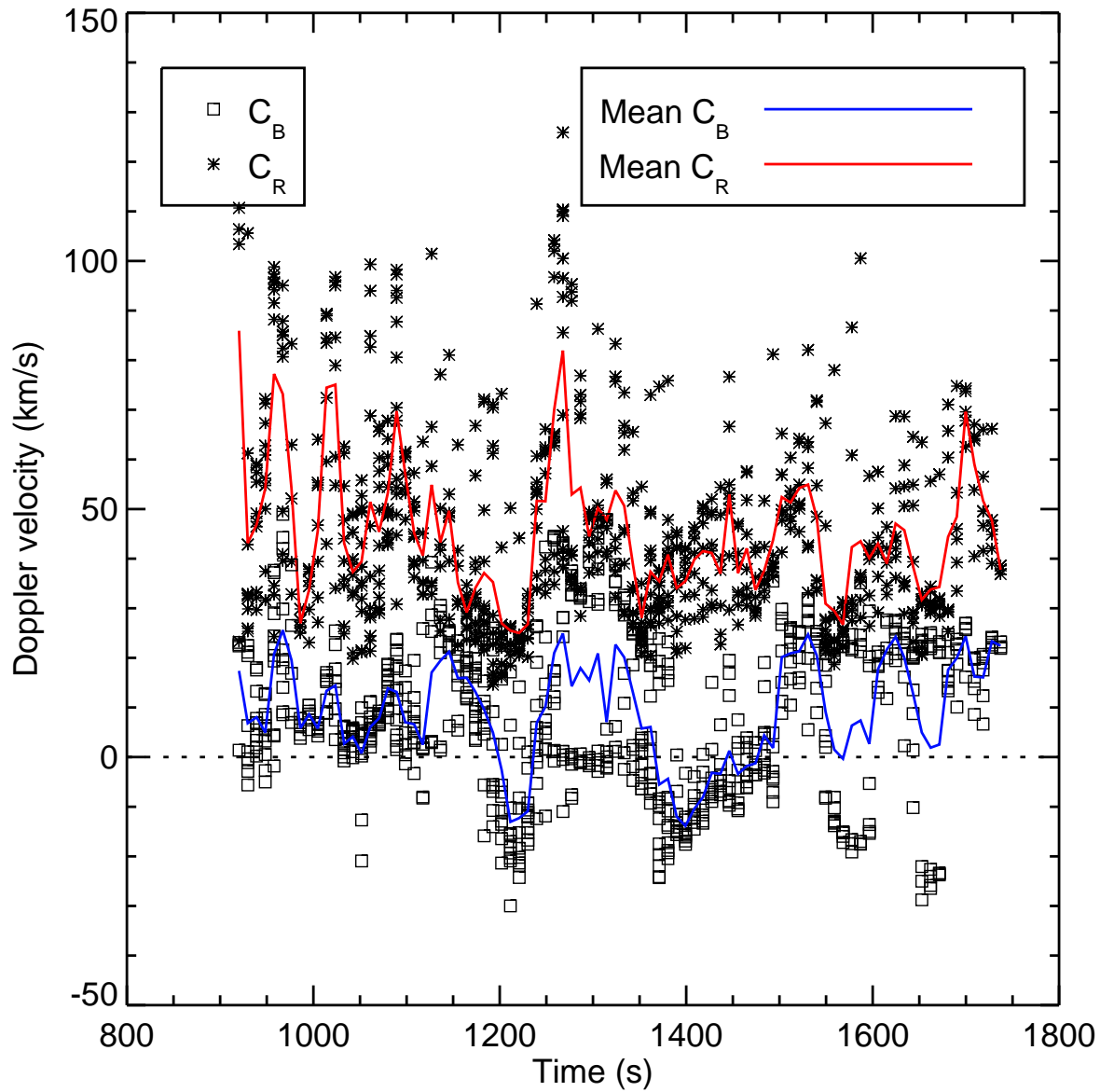}
\caption{Plot of the Doppler velocities (in km s$^{-1}$) for the two Si \textsc{iv} 1403 \AA\ components for all pixels within the sawtooth as a function of time (given in s after 12:33:38 UT).  The black squares (asterisks) are the Doppler velocities for component $C_B$ ($C_R$), and the solid blue (red) lines are the mean Doppler velocity for component $C_B$ ($C_R$) for all positions at a given time.
\label{fig:doppler_time}}
\end{figure}

A different method of visualizing the velocity evolution is to plot the Doppler velocities for all positions in a single time slice of the sawtooth as a function of time, as we have in Figure \ref{fig:doppler_time}.  Components $C_B$ and $C_R$ are displayed as the black squares and asterisks, respectively, and the mean velocity within each column is plotted as the solid blue (for $C_B$) and red (for $C_R$) lines.  In this format we can clearly observe that, starting at $\sim$1100 s, there begins a distinct red-blue oscillation in $C_B$, apparent in the mean velocity, which continues for at least three full periods of 100-200 s and amplitude $\pm$ 20 km s$^{-1}$.  Finally, we note that there is a similar oscillation in the redshifted $C_R$ Doppler velocity, which is noisier than for $C_B$ but still clearly at the same frequency as and in phase with the $C_B$ oscillation.  The similar frequency and phase of these Doppler oscillations strongly suggests that both Si \textsc{iv} components are being shifted by a common motion of the plasma.

\subsection{Additional spectral lines}
\label{sec:otherlines}

There are two additional pairs of spectral lines included in the IRIS SG that are of use for observing the upper chromosphere and TR.  One pair is the O \textsc{iv} 1400 \AA\ and 1401 \AA\ lines included in the FUV 2 passband.  These lines are formed higher in the TR than the Si \textsc{iv} lines, at $\sim$150,000 K, and as forbidden lines they are often of use for determining the plasma density at that temperature \citep{flower75}.  We find that the time-distance stackplot of the O \textsc{iv} 1401 \AA\ total line intensity is virtually indistinguishable in structure from that of Si \textsc{iv} 1403 \AA, with the sawtooth pattern appearing at an identical position and time.  Unlike the Si \textsc{iv} profiles discussed above, however, neither of the O \textsc{iv} lines have any obvious two-component structure and appear generally well-fit by a single Gaussian.  This may be because the O \textsc{iv} lines have count rates significantly lower than the Si \textsc{iv} lines, or it may be due to the specifics of the formation mechanism for the forbidden O \textsc{iv} lines.  Although we do not discuss it further in this work, we find that the single component O \textsc{iv} 1401 \AA\ Doppler shifts generally follow that of component $C_B$ of Si \textsc{iv} 1403 \AA, with similar magnitudes of blueshifts and redshifts located in the same parts of the sawtooth.

The other pair of lines are the C \textsc{ii} 1335 \AA\ and 1336 \AA\ lines in the FUV 1 passband.  These lines are formed in the upper chromosphere, at $\sim$20,000 K, and are thus expected to be (somewhat) closer to the footpoint than Si \textsc{iv} and generally will not display the same dynamics.  However, we find that C \textsc{ii} 1335 \AA\ and 1336 \AA\ lines also display nearly identical behavior to Si \textsc{iv}, with the sawtooth intensity pattern appearing over the same spacial and temporal scales. We also find that both C \textsc{ii} lines possess two-component Gaussian profiles, as was noted in \cite{cheng15}, and although we did not perform the same detailed analysis of the Doppler shifts we note that these components display very similar shifts to the Si \textsc{iv} lines at a several selected pixels within the sawtooth.

\subsection{The East ribbon}
\label{sec:east}

\begin{figure}
\epsscale{0.5}
\plotone{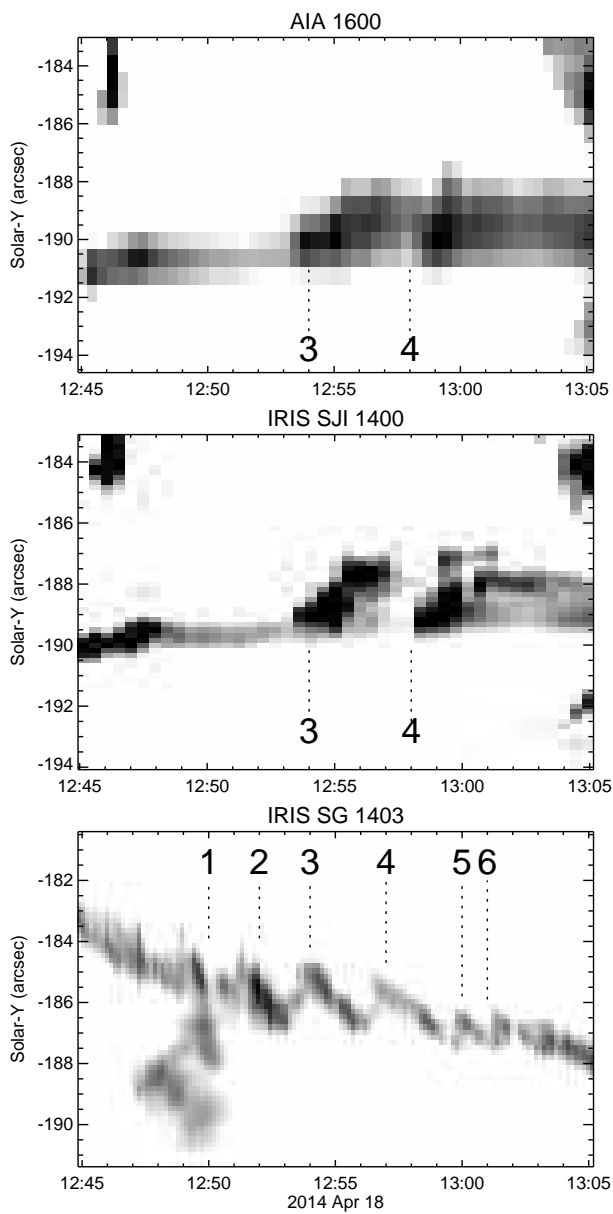}
\caption{Upper panel:  Time-distance stack plot of SDO/AIA 1600 \AA\ intensity for a vertical artificial slit across the ER (shown in Figures \ref{fig:aia1600} and \ref{fig:iris1400}), as described in Section \ref{sec:east}, with labels for the sawteeth ``3'' and ``4''.  Middle panel:  Time-distance stack plot of IRIS SJI 1400 \AA\ intensity for a vertical artificial slit across the ER (shown in Figures \ref{fig:aia1600} and \ref{fig:iris1400}), as described in Section \ref{sec:east}, with labels for the sawteeth ``3'' and ``4''.  Lower panel:  Reprint of the time-distance stackplot for total 1403 \AA\ intensity from Figure \ref{fig:intensity_stack} with sawtooth labels.  Time in all panels is in UT.
\label{fig:fake_slit}}
\end{figure}

Although the IRIS SG FOV only allows for spectral analysis of the WR, we note that the northern end of the ER was located with the IRIS SJI 1400 \AA\ FOV (Figure \ref{fig:iris1400}).  From the 171 \AA\ SDO/AIA image in Figure \ref{fig:aia171}, we observe a set of flare loops that emerge from the WR at the IRIS SG slit location which curve first North and then South and West to end at the ER at the location indicated in that figure by a vertical black ``I''-shaped line crossing the ER.  We extract an artificial SJI slit, one pixel wide and located along the vertical black ``I''-shaped lines in Figures \ref{fig:aia1600} \& \ref{fig:iris1400}, to investigate the ER intensity behavior in the same manner as the WR.  The resulting intensity time-distance stackplot is shown in the middle panel of Figure \ref{fig:fake_slit}, with the original SG stackplot from Figure \ref{fig:intensity_stack} plotted in the bottom panel.

We note immediately that there are two diagonal bands that begin at very nearly the same times as the peaks marked ``3'' and ``4'' in the original sawtooth.  Of course, the lower time resolution of the SJI ($\sim$27 s) means that any features in time are somewhat less clear than for the $\sim$9 s cadence SG.  However, the SJI bands both begin within 60 s of their companion SG sawtooth features, and their duration (120-180 s) and spatial extent (2-3$''$) are nearly identical.  We also note that the slow motion of the band is also directed away from the PIL, as it was for the SG sawtooth, indicating that the phase of these features is the same.  

We repeat this procedure using the co-aligned AIA 1600 data, which is often dominated by the C \textsc{iv} lines at 1548 \AA\ and 1550 \AA\ \citep{lemen12}. These lines are formed at $\sim$100,000 K, slightly hotter than for Si \textsc{iv}.  The resulting intensity stackplot is shown in the top panel of Figure \ref{fig:fake_slit}.  Once again, we note that the lower spatial ($0''.5$) and temporal (24 s) resolutions smear out the features, but we can still make out the same diagonal bands marked ``3'' and ``4'' at approximately the same positions and times as before.  Obviously, we cannot be absolutely certain that these ER footpoints are conjugate to the WR footpoints in the IRIS SG.  We also note that the bands present in the ER are not completely identical to the sawtooth in the WT, and that the oscillations ``1'', ``2'', ``5'', and ``6'' from the WR sawtooth are conspicuously missing in the ER.  Further, those oscillations do not appear for any other artificial slit positions on either side of the position shown, and as the artificial slit is moved away from that position the two oscillations that do appear are not as apparent.  This indicates some degree of asymmetry between the two ribbons.  However, we are fairly confidant stating that the conjugate behavior of the intensity stackplots indicates that both ribbons are experiencing the same oscillation, and that the close relationship we observe between the ER and WR strongly implicates a source somewhere high in the corona.  We therefore rule out any local mechanism, such as p-mode oscillations \citep{ting15}, which would not be linked by the coronal magnetic field.

\section{Interpretation}
\label{sec:interpretation}

\begin{figure}
\epsscale{0.5}
\plotone{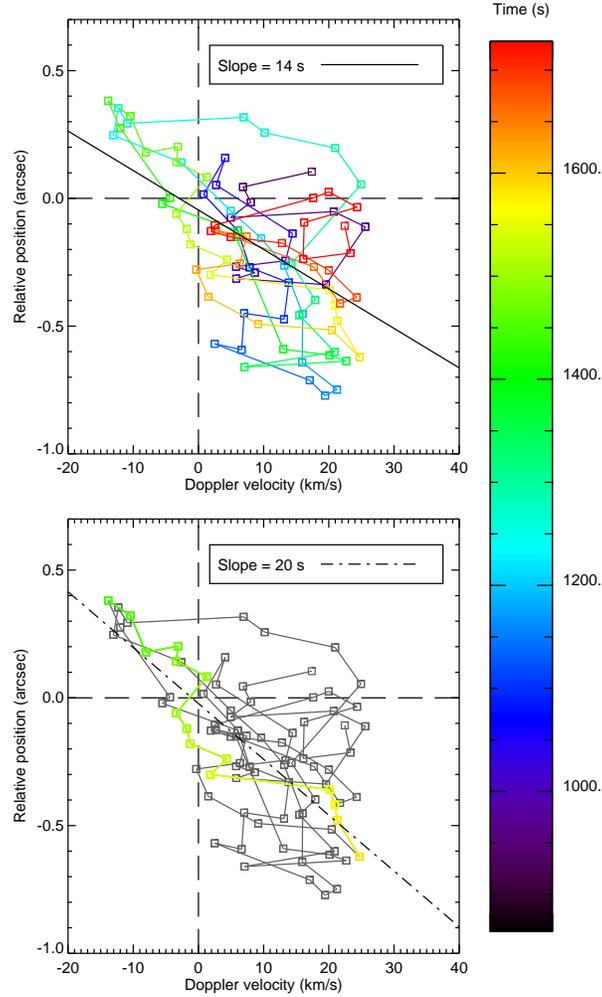}
\caption{Upper panel: phase portrait of the mean Doppler velocity of component $C_B$ (blue line from Figure \ref{fig:doppler_time}) and the relative sawtooth position (orange line subtracted from the blue line from Figure \ref{fig:intensity_stack}).  Data is indicated by the squares, and the connecting lines are to guide the reader.  Time is indicated by the color of the squares and connecting lines, and is indicated by the color bar at right.  The solid black line is a linear fit to all data.  Lower panel: same, but with only the data for sawtooth ``4'' colored (all other data in gray).  The dash-dotted line is a linear fit to the sawtooth ``4'' data points.
\label{fig:phase}}
\end{figure}

In the upper panel of Figure \ref{fig:phase} we have plotted the mean $C_B$ Doppler velocity in km s$^{-1}$, determined in Section \ref{sec:doppler} and shown in Figure \ref{fig:doppler_time}, against the relative difference in arcsec of the mean sawtooth position from the linear fit (effectively the orange line subtracted from the blue line in Figure \ref{fig:intensity_stack}).  The result is a phase portrait for the sawtooth ribbon oscillations.  Note that the actual sawtooth data are represented in the plot by the colored squares, with the connecting colored lines added to guide the reader.  Time is encoded by the color of the squares and connecting lines starting with purple at the beginning of the sawtooth and ending with red, as shown by the color bar on the right side of the figure.

\begin{figure}
\epsscale{1.0}
\plotone{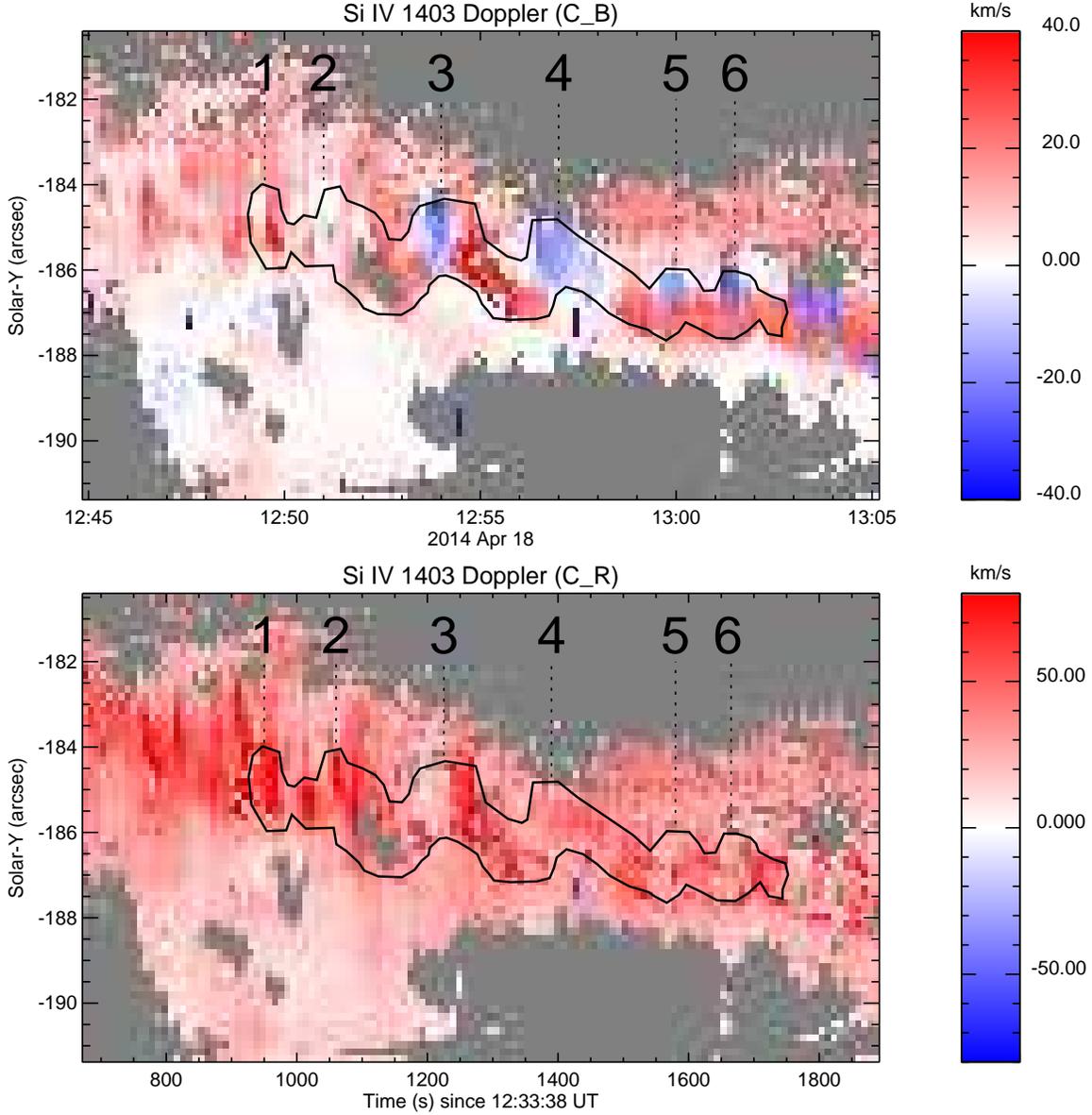}
\caption{Tme-distance stackplot of Doppler velocity for components $C_B$ (upper panel) and $C_R$ (lower panel) of the Si \textsc{iv} 1403 \AA spectral line.  Time is given on the $x$-axis in UT for the upper panel and in s after 12:33:38 UT for the lower panel.  The $y$-axis for both panels is solar-Y in arcsec.  Velocity scale is given to the right of each panel (scaled $\pm 40$ km s$^{-1}$ for $C_B$ and $\pm 80$ km s$^{-1}$ for $C_R$).  Black outline indicates the position of the sawtooth pattern, and gray pixels indicate discarded fits (as described in Section \ref{sec:fitting}).
\label{fig:doppler_stack}}
\end{figure}

The central feature of this phase portrait is that, despite some meandering of the data, many of the data appear to follow a diagonal trend that runs from upper left to lower right.  The solid black line represents a linear fit to the full data set within the sawtooth, and gives an overall slope of $\sim$14 s (using $\sim$900 km/arcsec).  However, we note that the diagonal trend becomes more apparent when we consider only individual pieces of the sawtooth.  In particular, in the lower panel of Figure \ref{fig:phase}, we have grayed-out all the data except that for sawtooth ``4'' in Figures \ref{fig:intensity_stack} and \ref{fig:doppler_stack} (beginning at $\sim$1400 s and ending at $\sim$1520 s).  These data are clearly aligned along a diagonal trend, and the dash-dotted line represents a linear fit to just these data with a slope of $\sim$20 s.  In the context of the discussion so far, the diagonal trend makes sense: the strongest blue and red Doppler shifts tend to be concentrated respectively at the peaks and troughs of the sawtooth pattern.  The phase portrait presented in Figure \ref{fig:phase} therefore reveals a critical property of the ribbon substructure.  Namely, that the LOS Doppler velocity and the position of the sawtooth pattern are $180^{\circ}$ out-of-phase.

Before presenting our proposed scenario for the generation of the sawtooth pattern, we first consider two alternative scenarios that would be probably serve as the most obvious first candidates.  The first scenario is a simple harmonic oscillation (SHO), where the loop is waving forward and back within a plane.  This could come from a linearly polarized loop oscillation and would create redshifts and blueshifts in the loop plasma as the loop motion carries it toward or away from the observer.  However, these Doppler shifts would occur only as the loop was moving from North to South or from South to North, and cannot account for the Doppler shifts at the peaks and troughs of the wave.  We also note that the phase portrait for an SHO is an ellipse in position-velocity space, resulting from the $90^{\circ}$ phase difference between position and velocity.  The SHO scenario is therefore inconsistent with the results from the ribbon sawtooth.

The second scenario which we consider is also an SHO ribbon oscillation, but instead of the loop motion creating the Doppler shifts it simply modulates pre-existing plasma velocities within the loop.  In this scenario, the loop tilts first toward and then away from the observer, the Doppler shifts of plasma flowing along the loop would be changed from red to blue (or vice versa, depending on the flow direction).  Assuming the correct conditions, this modulation could provide the correct $180^{\circ}$ phase difference.  It is readily seen, however, that this scenario requires the loop oscillation to be at least partly on the far side of $90^{\circ}$ inclination from the LOS, in order to switch the sign of the Doppler shift.  From the 45-minute HMI vector magnetogram at 12:58:19 UT we find that the magnetic field in the sawtooth region is angled only $\sim$$40^{\circ}$ from the LOS.  This would require a substantial ($>$$50^{\circ}$) oscillation in order to switch the Doppler shifts, which does not seem reasonable given the observations.

The $180^{\circ}$ phase difference between the LOS velocity and the apparent wave position suggests that the flare loops (and the fluid elements of plasma along them) are undergoing motion that is alternately parallel and perpendicular to the LOS.  One example of a class of waves which produces such motion are surface waves, such as occurs in deep water driven by wind.  In a surface wave, traveling in a direction $x$ perpendicular to the surface normal $z$ with no mean flow, the displacement of a fluid element $\mathbf{\delta x}$ is given by
\begin{equation}
\mathbf{\delta x} \left( x,z,t \right) = A \textrm{e}^{kz} \left[ -\sin \left( kx-\omega t \right) \mathbf{\hat{e}_{x}} + \cos \left( kx-\omega t \right) \mathbf{\hat{e}_{z}} \right];
\end{equation}
and the the velocity $\mathbf{v}$ is given by
\begin{equation}
\mathbf{v} \left( x,z,t \right) = A \omega \textrm{e}^{kz} \left[ \cos \left( kx-\omega t \right) \mathbf{\hat{e}_{x}} + \sin \left( kx-\omega t \right) \mathbf{\hat{e}_{z}} \right],
\end{equation}
where $A$ is the wave amplitude, $\omega$ is the wave frequency, $k$ is the wave number, and $\mathbf{\hat{e}_{j}}$ is the $j^{{\scriptsize\textrm{th}}}$ unit vector \citep{phillips77}.  As can be seen from these equations, a fluid element in this wave will trace out an ellipse in $\left( x,z \right)$ as the wave passes, with a $90^{\circ}$ phase difference between the two velocity components and a total $180^{\circ}$ phase difference between a given velocity component and the position component perpendicular to it.

We also recall that the symmetry between the two ribbons suggests a coronal source to the wave.  It therefore seems likely that an instability in the current sheet during reconnection is driving the sawtooth.  Two such instabilities which have been shown to occur in coronal current sheets during reconnection, and which can lead to oscillations in the subsequent flare loops, are the {\em Kelvin-Helmholtz} (KH) instability \citep{uchimoto91,ofman11,foullon11} and the {\em tearing-mode} (TM) instability \citep{furth63}.  Like the water wave described above, both of these plasma instabilities are confined to a surface (the current sheet) and produce elliptical motions.  In the upper panel of Figure \ref{fig:diagram}, we present a schematic conception of the scenario by which these instabilities could produce the observed features.  Looking north along the flare arcade, the LOS comes roughly from the upper left.  The outer two black lines represent a pair of reconnecting field lines, where the reconnection is occurring at the star along the current sheet (vertical dashed line).  We then hypothesize that either a KH or TM instability occurs, which results in elliptical motion of the field lines during their reconnection (indicated by the arrowed circles).  The oscillation frequency is low enough that the loop responds rigidly, as a whole, pivoting at some depth below the chromosphere.  Finally, the inner black line shows a previously reconnected loop that has contracted and then cooled to become visible in 171 \AA.  By this point the loop is disconnected from the current sheet and oscillation has ceased.  The instability drives the elliptical oscillation, but also appears to have a phase velocity within the sheet, producing the pattern motion at $v_{st}\simeq15$ km s$^{-1}$.

Meanwhile, at the loop footpoint we see the effects of the instability-driven oscillations on the ribbon.  The dotted box in the upper panel of Figure \ref{fig:diagram} indicates the zoom-in region, which is shown in the lower panel of Figure \ref{fig:diagram}.  Here, the oscillating loop is shown in two positions as the solid arrowed lines, with the LOS again from upper left as indicated.  The loop is shown anchored somewhere deeper in the atmosphere, and the elliptical loop oscillation is indicated by the ellipse at the center and by the $\otimes$ and $\odot$ respectively representing motion into and out of the page.  The observed amplitude of the oscillation will be the diameter of the ellipse; from above we find this to be $\sim$1-3$''$ or $\sim$0.7-2.2 Mm.  We note that the only motion of the loop that results in Doppler shifts is that of the loop moving along the LOS, as indicated by the blue and red arrows, and that the loop motion at the $\otimes$ and $\odot$ is perpendicular to the LOS and generates no Doppler shifts.  Therefore, as we observe, the blue and red Doppler shifts will occur at the northern and southern extremes of the oscillation, as in Figure \ref{fig:doppler_stack}, and that the phase difference between the apparent position and Doppler shift will be $180^{\circ}$.

Obviously, in order to produce the observed Doppler shifts in $C_B$ of $\pm$20 km s$^{-1}$, the loop must be rotating around the circle indicated in Figure \ref{fig:diagram} at approximately that same speed (technically it would need to be somewhat faster, due to the inclined LOS).  Given that the observed amplitude of the oscillation (equivalent to the diameter of the circle) is $\sim$0.7-2.2 Mm, we note that a rotation speed of $\sim$30 km s$^{-1}$ (allowing for LOS effects) would give an oscillation period of 70-230 s.  This is very similar to the observed range for the period of the sawtooth oscillation, which varied between 80-190 s.

Finally, we believe that the two components $C_R$ and $C_B$ occur in distinct sets of loop footpoints, which are too closely spaced to be resolved by IRIS.  Both sets of loops are participating in the elliptical oscillation generated by the instability described above.  However, the greater non-thermal broadening of $C_R$, as well as the nonzero average redshifts, suggests that those footpoints are undergoing chromospheric condensation, presumably as that plasma is being directly energized by reconnection in the instability region.  The $C_B$ component, on the other hand, has non-thermal broadening consistent with previous quiescent AR observations \citep{depontieu15}, and displays an approximately zero average Doppler shift during the oscillation.  This suggests that this set of footpoints have not been energized into condensation by coronal reconnection, even though those loops are still being elliptically oscillated by the instability.  We recall, however, that the $C_B$ component is still significantly brightened during the oscillation; it is possible that this plasma is being energized into enhanced emission via energy transfer from the loop waves \citep{russell13}.

\section{Discussion}
\label{sec:discussion}

In this paper, we have presented an analysis of a two-ribbon flare using IRIS and AIA imaging and spectral observations.  We found, along with the usual ribbon spreading at $\sim$1-2 km s$^{-1}$, that the ribbons show a distinct sawtooth substructure with a scale of 1-2 Mm.  The sawtooth appears in both ribbons, maintains its shape coherently over time, and drifts along the ribbon from east to west at a speed of $v_{st}\simeq15$ km s$^{-1}$.  The period of this sawtooth oscillation was found to average $\sim$140 s, and the oscillation speed is $\sim$20 km s$^{-1}$.  We also found that the sawtooth substructure appears in a variety of spectral lines, including Si \textsc{iv}, O \textsc{iv}, and C \textsc{ii}, with nearly identical amplitude and phase across a wide range of temperatures.  We observed that the line profiles for Si \textsc{iv} and C \textsc{ii} in and around the sawtooth have two major components (previously noted by \cite{cheng15}), and we developed an automated routine for fitting the two Si \textsc{iv} lines observed by the IRIS SG.  O \textsc{iv} does not clearly show two components, but we noted that this may be due to the weaker intensity or different formation mechanism of the forbidden O \textsc{iv} lines.  We showed that the two components of Si \textsc{iv} tend to persist over time and we identified the two components by their relative Doppler shifts, with the redder component labeled $C_R$ and the bluer component $C_B$.

We found that $C_R$ was redshifted by an average $\sim$50 km s$^{-1}$ during the sawtooth, whereas $C_B$ switched between redshifts and blueshifts.  Further, we noted that the Doppler shifts for both components oscillate during the sawtooth with an amplitude of $\pm$20 km s$^{-1}$ (averaging $\sim$50 km s$^{-1}$ for $C_R$ and $\sim$0 km s$^{-1}$ for $C_B$) and a period of 100-200 s.  This Doppler oscillation was found to be correlated with the spatial sawtooth oscillation, and the phase between these was found to be $180^{\circ}$.  We also used the flare loops, identified in SDO/AIA 171 \AA, to find the approximate conjugate point in the ER.  Using an artificial slit constructed from the IRIS SJI 1400 \AA\ and SDO/AIA 1600 \AA\ data, we were able to identify two bands with the same amplitude, period, and phase as the WR sawtooth substructure.  We believe the conjugacy of the sawtooth oscillation (along with the continuous and coherent evolution of the sawtooth) strongly suggests that the sawtooth is {\em not} the result of field-aligned transport effects, and that the two ribbons are being influenced by a common coronal source for the oscillation. We use $180^{\circ}$ phase difference between the Doppler shift and sawtooth position, as well as the LOS inclination of the magnetic field, to eliminate two simple wave modes as the source for the sawtooth.  Finally, we propose an elliptical oscillation motion of the loop, driven by an instability in the coronal current sheet, as the mechanism behind the observed sawtooth ribbon substructure.

\begin{figure}
\epsscale{1.0}
\plotone{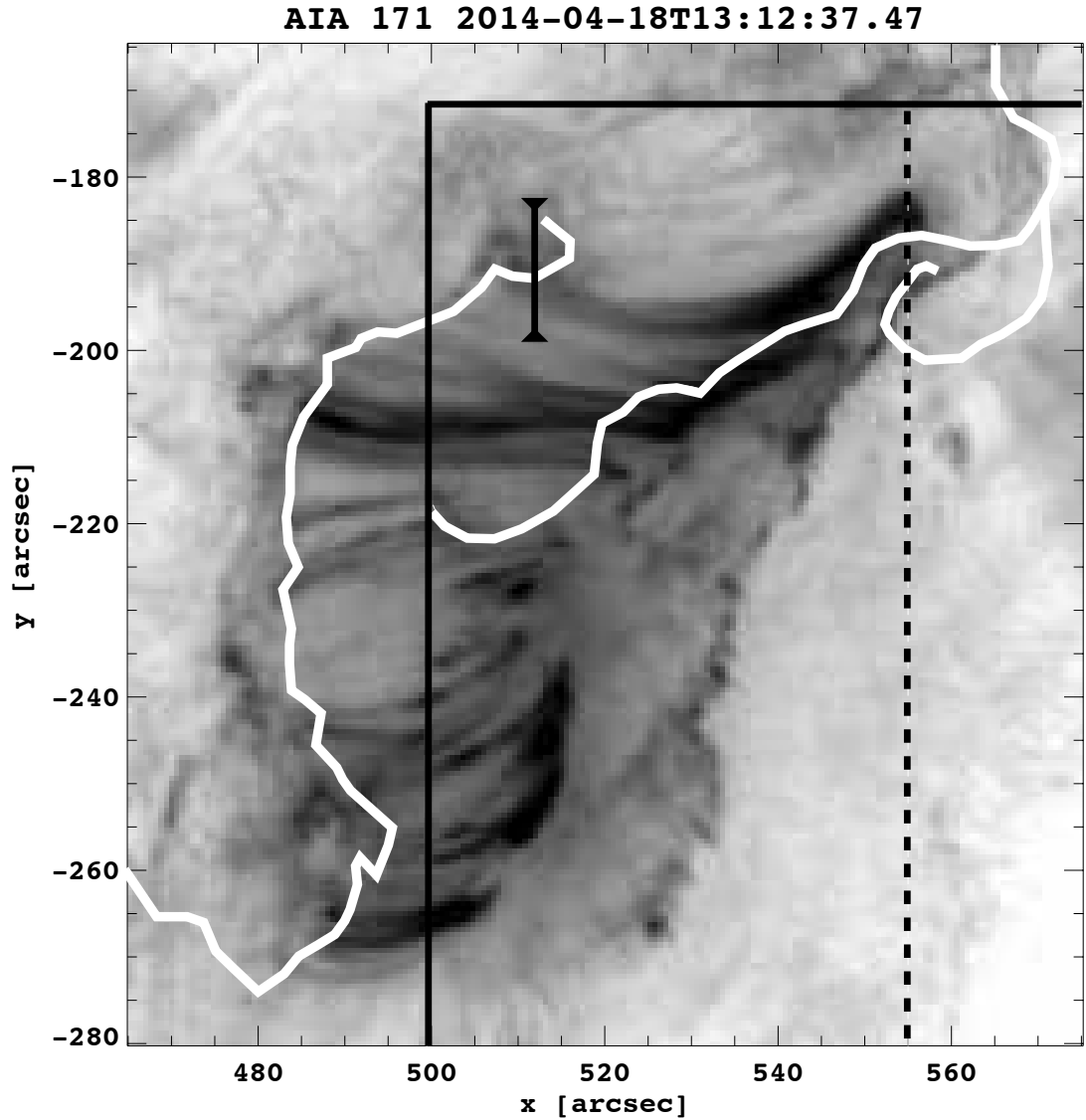}
\caption{SDO/AIA 171 \AA\ image of the post-flare loops and ribbons, in RLBW.  The black inset box is the positioning of the IRIS SJI FOV, and the vertical black dashed line is the IRIS SG slit position.  The solid white lines trace the position of the SDO/AIA 1600 ribbons from Figure \ref{fig:aia1600}, and the black ``I''-shaped line is an artificial slit discussed in Section \ref{sec:east}.
\label{fig:aia171}}
\end{figure}

\begin{figure}
\epsscale{0.5}
\plotone{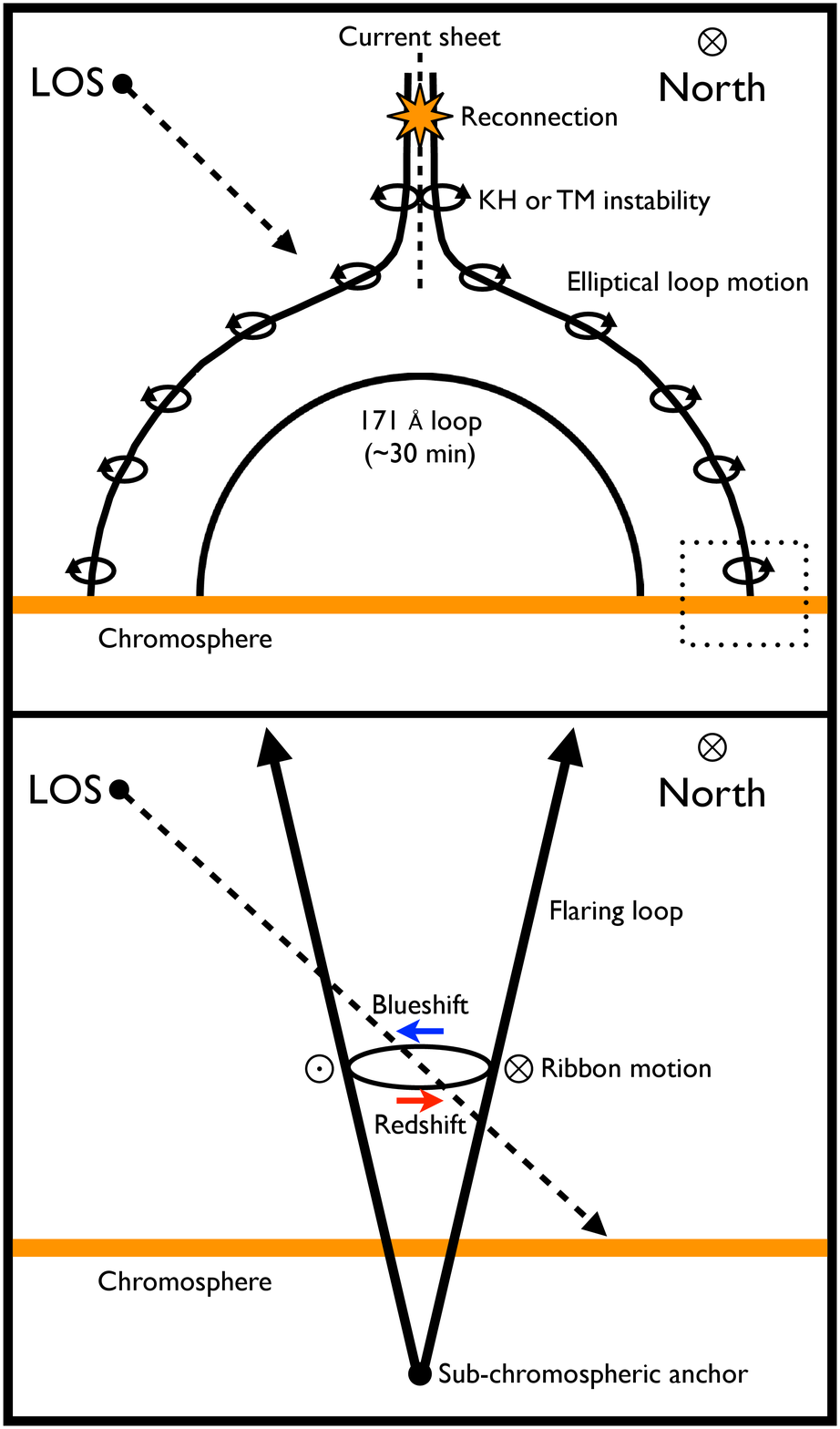}
\caption{Upper panel: schematic diagram of our proposed scenario, with a KH or TM instability in the coronal current sheet resulting in elliptical wave oscillations in the reconnected flare loops.  Lower panel: schematic diagram of the flare loop footpoint, as described in Section \ref{sec:interpretation}, showing how the elliptical wave motion relative to the LOS generates the observed $180^{\circ}$ phase difference between the ribbon motion and Doppler velocities.
\label{fig:diagram}}
\end{figure}

The scenario depicted in Figure \ref{fig:diagram} is reminiscent of the kinds of loop oscillations sometimes observed in flares \citep{aschwanden99}.  Oscillation of loops in 171 \AA\ have previously been observed in association with KH instabilities \citep{ofman11}.  We failed to observe any oscillations in the AIA 171 \AA\ image sequence which includes Figure \ref{fig:aia171}, but this may be due to the $\sim$30 min delay between the loop formation, by reconnection, and its appearance at that cool wavelength.

Oscillations triggered by flares are interpreted as standing MHD waves, and can have periods in the $\sim$200 s range, consistent with the Alfv\'en transit time in the long, high, weak loops which are often found to oscillate.  This interpretation is more problematic for our case with  short, low, strong loops.  The loops, visible in 171 \AA\ (see Figure \ref{fig:aia171}) appear to fit field lines of a constant-alpha field with $\alpha\simeq-5\times10^{-9}\,{\rm m}^{-1}$, extrapolated form the line-of-sight HMI magnetogram.  These field lines range in length from $L=40$ to $60$ Mm, and have field strengths falling to $B\sim 150$ G  at their apices, around $z\sim10$ Mm.  For a typical density of $n_e\simeq10^9\,{\rm cm}$, the coronal Alfv\'en speed would be no smaller than 10 Mm s$^{-1}$, for which the end-to-end transit times would be 4--6 seconds on these loops.  The 140 s sawtooth evident at the ribbon could not, therefore, correspond to a standing wave in loops like those formed after the flare.

We believe a more likely scenario to be that the short, strong field lines are moving quasi-statically (far slower than the Alfv\'en speed) in response to motion imposed by an elliptical wave occurring at or near the current sheet.  This could be the hydrodynamic motion of a KH instability driven by velocity shear across the current sheet \citep{ofman11}.  For this to be the case, however, the magnetic field would have to be very nearly anti-parallel across the sheet, in order that a guide-field component not stabilize the instability.  If the pattern speed, $v_{st}\sim 15$ km/s were related to this component of the Alfv\'en speed the angle between the fields would have to be within one milliradian ($0.006^{\circ}$) of perfectly anti-parallel.  A more likely explanation is the tearing mode (TM), which is expected to occur at current sheets, and also exhibits elliptical flow patterns (i.e.\ the velocity stream function consists of islands).  The observed ribbon motion would be related to the inflows and outflows from the X-lines, estimated to be smaller than the Alfv\'en speed by a factor, $S^{-1/2}$, where $S\gg1$ is the Lundquist number of the current sheet.  Flow speeds of $\sim 20$ km/sec would arise from a Lundquist number, $S\sim 10^6$, but this may reflect the turbulent state of the sheet itself.  Finally, this instability would need to propagate at a speed lower than Alfv\'en speed, but comparable to the flow speeds of the instability.  However, we have no insights as to what might give the instability the precise phase velocity that we observe.

Another point concerning the proposed scenario in Figure \ref{fig:diagram} which we left unspecified is the location of the loop anchor.  Presumably, this anchor point is located deep enough in the atmosphere to allow for the observed sawtooth amplitude.  We might expect that the Si \textsc{iv}, O \textsc{iv}, and C \textsc{ii} spectral lines , which are normally formed at different heights within the chromosphere and TR and would thus exhibit different amplitudes of the loop motion, providing us with some information about the anchor location.  However, as we noted, these lines instead all display identical amplitudes of the sawtooth pattern.  We speculate that this is due to the TR being compressed during the flare, such that the normally separated spectral line formation heights are all approximately the same within the ribbon, which eliminates any information about the location of the loop anchor.

We also note that the small mean Doppler shift of $C_B$ indicate that the plasma generating that component of the Si \textsc{iv} emission is not probably undergoing intense evaporation during the sawtooth.  Further, the non-thermal widths for $C_B$ are consistent with those for quiescent AR plasma \citep{depontieu15}.  On the other hand, both components of Si \textsc{iv} show strong increases in intensity during the sawtooth, indicating the the plasma has been energized into enhanced emission.  We do not know what is causing this brightening of non-evaporating plasma within the context of our proposed scenario, although we have speculated above that wave motion may contribute to the plasma energization.

In their recent work, \cite{ting15} report an observation of ribbon substructure in a different flare (2014 September 10, X1.6-class)  that is in some ways very similar to our observations of the 2014 April 18 flare.  In particular, they observed a quasi-periodic slipping motion, which manifested as a series of bright knots which appear to move along the ribbon at $\sim$ 20 km s$^{-1}$, close to our own estimate of $v_{st}$.  Their interpretation of this slipping motion was as an apparent motion, of the kind discussed by \cite{aulanier06} and dubbed ``slipping reconnection''.  Unlike our scenario for the 2014 April 18 event, however, they speculate that this slipping reconnection may be the result of density variations in the reconnection region, which are driven by p-mode oscillations above the sunspot.  Also in marked contrast to our observations, \cite{ting15} report a relatively steady Si \textsc{iv} redshift, unmodulated by the sawtooth pattern.  This is consistent with an interpretation a pattern projected from the current sheet onto the chromosphere along stationary magnetic field lines: i.e.\ slipping reconnection.  Our observations, however, reveal clear motion of the plasma, through the Doppler shifts which are correlated with the sawtooth pattern.  This rules out the hypothesis that the oscillations are merely projections and, at least in our flare, it rules out slipping reconnection.

It is possible that the discrepancies between these two observations arise purely from the different viewing angles of the different cases.  The 2014 September 10 flare occurred very close to disk center, so if there were horizontal plasma motions, tracking the elliptical loop oscillations, they would not give rise to line-of-sight Doppler shifts.  Our own case, the 2014 April 18 flare, occurred 40 degrees from disk center, where horizontal motions would, and apparently do, produce line-of-sight Doppler shifts.

The apparently contradictory observations presented in this paper and in \cite{ting15} indicate that additional observations of flare ribbons, located at varying locations and viewing angles, will be necessary to differentiate between the two hypothesis: apparent motion due to slipping reconnection {\em vs} elliptical loop oscillations driven by KH or TM instabilities.  Of course, it has only been recently that instruments with sufficient spatial and temporal resolution have allowed for observations of ribbon substructure.  As we have noted, lower resolution and cadence imagers such as SDO/AIA only barely allow for recognizing ribbon substructure at scales less than $1''$, and also do not provide the Doppler information that allows for determining phase relationships of chromospheric phenomena.  The authors expect that as future IRIS observations develop, particularly in conjunction with a coronal spectrograph such as {\em Hinode}/EIS, we will be able to distinguish between various hypothesis for generating ribbon substructure, and use observations of ribbons to eliminate or support models for reconnection in the corona.

\acknowledgments

IRIS is a NASA small explorer mission developed and operated by LMSAL with mission operations executed at NASA Ames Research center and major contributions to downlink communications funded by the Norwegian Space Center (NSC, Norway) through an ESA PRODEX contract.  The authors would like to thank the anonymous referee who provided suggestions for significant improvement of the original manuscript.  The authors would also like to thank Prof.\ Charles Kankelborg, Dr.\ Sarah Jaeggli, Dr.\ Ying Li, Dr.\ Paola Testa, and Dr.\ Bart De Pontieu for helpful and productive discussions concerning the IRIS instruments and data analysis.  This work was supported by contract 8100002702 from Lockheed Martin to Montana State University, a Montana Space Grant Consortium graduate student fellowship, and by NASA through HSR.

\clearpage



\clearpage

\clearpage

\clearpage

\clearpage

\clearpage

\clearpage

\clearpage

\clearpage

\clearpage

\clearpage

\clearpage

\clearpage



\begin{thebibliography}{}


\bibitem[Aschwanden et al.(1999)]{aschwanden99} Aschwanden, M.J., Fletcher, L., Schrijver, C.J., et al. 1999 ApJ 520, 880
\bibitem[Aulanier et al.(2006)]{aulanier06} Aulanier, G., Pariat, E., D\'emoulin, P., et al. 2006, SoPh 238, 347
\bibitem[Brosius \& Phillips(2004)]{brosius04} Brosius, J. W., \& Phillips, K. J. H. 2004, ApJ 613, 580
\bibitem[Brown(1973)]{brown73} Brown, J.C. 1973, SoPh 31, 143
\bibitem[Cheng et al.(2015)]{cheng15} Cheng, X., Ding, M.D., \& Fang, C. 2015, submitted to ApJ
\bibitem[Craig \& McClymont(1976)]{craig76} Craig, I.J.D. \& McClymont, A.N. 1976, SoPh 50, 133
\bibitem[De Pontieu et al.(2014)]{iris14} De Pontieu, B., Title, A.M., Lemen, J.R., et al. 2014, SoPh 289, 2733
\bibitem[De Pontieu et al.(2015)]{depontieu15} De Pontieu, B., McIntosh, S., Martinex-Sykora, J., et al. 2015, ApJL 799, L12
\bibitem[Del Zanna et al.(2002)]{delzanna02} Del Zanna, G., Landini, M., \& Mason, H.E. 2002, A\&A 385, 968
\bibitem[Falchi et al.(1997)]{falchi97} Falchi, A., Qiu, J., \& Cauzzi, G. 1997, A\&A 328, 371
\bibitem[Fletcher \& Warren(2003)]{fletcher03} Fletcher, L., \& Warren, H. P. 2003, `Energy Conversion and Particle Acceleration in the Solar Corona', L. Klein (ed.), Springer Lecture Notes in Physics 612, 58
\bibitem[Fletcher et al.(2004)]{fletcher04} Fletcher, L., Pollock, J.A., \& Potts, H.E. 2004, SoPh 222, 279
\bibitem[Flower \& Nussbaumer(1975)]{flower75} Flower, D.R., \& Nussbaumer, H. 1975, A\&A 45, 145
\bibitem[Forbes \& Priest(1984)]{forbes84} Forbes, T. G., \& Priest, E. R. 1984, SoPh 94, 315
\bibitem[Forbes et al.(1989)]{forbes89} Forbes, T.G., Malherbe, J.M., Priest, E.R. 1989, SoPh 120, 285
\bibitem[Foullon et al.(2011)]{foullon11} Foullon, C., Verwichte, E., Nakariakov, V. M., et al. 2011, ApJL 729, L8
\bibitem[Furth et al.(1963)]{furth63} Furth, H.P., Killeen, J., \& Rosenbluth, M.N. 1963, PhFl 6, 459
\bibitem[Handy et al.(1999)]{handy99} Handy, B. N., Acton, L. W., Kankelborg, C. C., et al. 1999, SoPh 187, 229
\bibitem[Ichimoto \& Kurokawa(1984)]{ichimoto84} Ichimoto, K., \& Kurokawa, H. 1984, SoPh 93, 105
\bibitem[Isobe et al.(2005)]{isobe05} Isobe, H., Takasaki, H., \& Shibata, K. 2005, ApJ 632, 1184
\bibitem[IRIS Technical Note 20()]{itn20} Tian, H., Wuelser, J. P., Boerner, P., et al. 2014, ``IRIS Technical Note 20: Wavelength Calibration''
\bibitem[Karpen et al.(1995)]{karpen95} Karpen, J.T., Antiochos, S.K., \& DeVore, C..R. 1995, ApJ 450, 422
\bibitem[Kerr et al.(2005)]{kerr05} Kerr, F.M., Rose, S.J., Wark, J.S., et al. 2005, ApJ 629, 1091
\bibitem[Kliem et al.(2000)]{kliem00} Kliem, B., Karlick\'y, M., \& Benz, A.O. 2000, A\&A 360, 715
\bibitem[Kopp \& Pneumann(1976)]{kopp76} Kopp, R. A., \& Pneuman, G. W. 1976, SoPh 50, 85
\bibitem[Lemen et al.(2012)]{lemen12} Lemen, J.R., Title, A.M., Akin, D.J., et al. 2012, SoPh 275, 17
\bibitem[Li \& Zhang(2015)]{ting15} Li, T., \& Zhang, J. 2015, accepted to ApJ
\bibitem[Longcope et al.(2007)]{longcope07} Longcope, D., Beveridge, C., Qiu, J., et al. 2007 SoPh 244, 45
\bibitem[Longcope et al.(2009)]{longcope09} Longcope, D.W., Guidoni, S.E., \& Linton, M.G. 2009 ApJ 690, L18
\bibitem[Lysak \& Song(1996)]{lysak96} Lysak, R.L., \& Song, Y. 1996, Journal of Geophysical Research 101, 15411
\bibitem[Mathioudakis et al.(1999)]{mathioudakis99} Mathioudakis, M., McKenny, J., Keenan, F.P., et al. 1999, A\&A Letter 351, L23
\bibitem[Miklenic et al.(2007)]{miklenic07} Miklenic, C. H., Veronig, A. M., Vr\u{s}nak, B., et al. 2007 A\&A 461, 697
\bibitem[Milligan \& Dennis(2009)]{milligan09} Milligan, R. O., \& Dennis, B. R. 2009, ApJ 699, 968
\bibitem[Nishizuka et al.(2009)]{nishizuka09} Nishizuka, N., Asai, A., Takasaki, H., et al. 2009, ApJL 694, L74
\bibitem[Ofman \& Thompson(2011)]{ofman11} Ofman, L., \& Thompson, B.J. 1982, ApJL 734, L11
\bibitem[Phillips(1977)]{phillips77} Phillips, O.M. 1977, The Dynamics of the Upper Ocean (Cambridge: Cambridge Univ. Press)
\bibitem[Priest(1985)]{priest85} Priest, E.R. 1985 Rep. Prog. Phys. 48, 955
\bibitem[Qiu et al.(2002)]{qiu02} Qiu, J., Lee, J., Gary, D.E., et al. 2002, ApJ 565, 1335
\bibitem[Qiu(2009)]{qiu09} Qiu, J. 2009 ApJ 692, 1110
\bibitem[Russell \& Fletcher(2013)]{russell13} Russell, A. J. B., \& Fletcher, L. 2013, ApJ 765, 81
\bibitem[Schmieder et al.(1987)]{schmieder87} Schmieder, B., Forbes, T. G., Malherbe, J. M., et al. 1987 ApJ 317, 956
\bibitem[Shibata \& Tanuma(2001)]{shibata01} Shibata, K., \& Tanuma, S. 2001, Earth, Planets and Space 53, 473
\bibitem[Uchimoto et al.(1991)]{uchimoto91} Uchimoto, E., Strauss, H.R., \& Lawson, W.S. 1991, SoPh 134, 111
\bibitem[Warren \& Warshall(2001)]{warren01} Warren, H.P., \& Warshall, A.D. 2001 ApJ 560, L87

\end{thebibliography}
\end{document}